\documentclass[prx,aps,amssymb,twocolumn,reprint,superscriptaddress,longbibliography]{revtex4-1}
\usepackage{amsmath}
\usepackage{amssymb}
\usepackage{amsthm}
\usepackage{amsfonts}
\usepackage{listings}
\usepackage{enumerate}
\usepackage{latexsym}
\usepackage{color}
\usepackage{xcolor}
\usepackage{bm}
\usepackage{hyperref}
\hypersetup{
 pdfnewwindow=true, colorlinks=true,
 linkcolor=blue, anchorcolor=blue,
 citecolor=blue, filecolor=blue,
 menucolor=blue, urlcolor=blue}

\newcommand{\CSS}{${\rm Co_3Sn_2S_2}$}
\usepackage{psfrag}

\usepackage{graphicx}
\usepackage{subfigure}
\usepackage{lipsum}
\usepackage{float}
\usepackage{array} 
\usepackage{soul}

\newcommand*{\dif}{\mathop{}\!\mathrm{d}}

\begin{document}
\tolerance 10000

\draft

\title{A combined First-principles and Boltzmann transport theory methodology for studying  magnetotransport in magnetic materials}
\author{Zhihao Liu}
\affiliation{Beijing National Laboratory for Condensed Matter Physics,
and Institute of Physics, Chinese Academy of Sciences, Beijing 100190, China}
\affiliation{University of Chinese Academy of Sciences, Beijing 100049, China}

\author{Shengnan Zhang}
\affiliation{Beijing National Laboratory for Condensed Matter Physics,
and Institute of Physics, Chinese Academy of Sciences, Beijing 100190, China}
\author{Zhong Fang}
\affiliation{Beijing National Laboratory for Condensed Matter Physics,
and Institute of Physics, Chinese Academy of Sciences, Beijing 100190, China}
\affiliation{University of Chinese Academy of Sciences, Beijing 100049, China}

\author{Hongming Weng}
\email{hmweng@iphy.ac.cn}
\affiliation{Beijing National Laboratory for Condensed Matter Physics,
and Institute of Physics, Chinese Academy of Sciences, Beijing 100190, China}
\affiliation{University of Chinese Academy of Sciences, Beijing 100049, China}
\affiliation{Songshan Lake Materials Laboratory, Dongguan, Guangdong 523808, China}

\author{Quansheng Wu}
\email{quansheng.wu@iphy.ac.cn}
\affiliation{Beijing National Laboratory for Condensed Matter Physics,
and Institute of Physics, Chinese Academy of Sciences, Beijing 100190, China}
\affiliation{University of Chinese Academy of Sciences, Beijing 100049, China}

\begin{abstract}
Unusual magnetotransport behaviors, such as temperature-dependent negative magnetoresistance (MR) and bowtie-shaped MR, have puzzled us for a long time. Although several mechanisms have been proposed to explain these phenomena, the absence of comprehensive quantitative calculations has made these explanations less convincing. In our work, we introduce a methodology to study magnetotransport behaviors in magnetic materials. This approach integrates the anomalous Hall conductivity induced by the Berry curvature, with a multi-band ordinary conductivity tensor, employing a combination of first-principles calculations and semi-classical Boltzmann transport theory. Our method also incorporates the temperature dependence of the relaxation time and the anomalous Hall conductivity, as well as the field dependence of the anomalous Hall conductivity. We initially test this approach on models, obtaining distinct behaviors of magnetoresistance and Hall resistivity across several types of magnetic materials, and then apply it to a Weyl semimetal, \CSS. The results, which align well with experimental observations in terms of magnetic field and temperature dependencies, demonstrate the efficacy of our approach. This methodology provides a comprehensive and efficient means to understand the underlying mechanisms of the unusual complex behaviors observed in magnetotransport measurements in magnetic materials.
\end{abstract}

\maketitle
\section{Introduction}
Large anomalous Hall effect (AHE)~\cite{shekhar2018anomalous,wang2018large_nature.commu,li2020giant,PhysRevLett.126.106601_giant,fujishiro2021giant,LiMnSn_2021,GdAuGe_2023} and unusual magnetotransport behaviors such as large magnetoresistance (MR)~\cite{ali2014large_WTe2,shekhar2015extremely_NbP,kumar2017extremely_WP2,GdAuGe_2023} or negative magnetoresistance (NMR)~\cite{doi:10.1021/acs.nanolett.0c02219, BaMnBi_2021, Mn3Sn_2022, TbPdBi_2023} are often observed in magnetic materials. Exploration of various MR and large AHE responses may lead to novel electronic functionalities and efficient spintronic devices. Among them, the most intriguing phenomenon is the NMR. Several mechanisms have been proposed in numerous literature to explain the NMR, such as weak localization~\cite{BERGMANN19841weaklocalization}, electron-magnon scattering~\cite{mihai2008electron-magnon_prb,raquet2002electron-magnon_prb,zhao2023magnetoresistance}, etc. However, few studies quantitatively investigate the impact of the anomalous Hall effect on MR, especially the NMR, in magnetic materials. Actually, the richness and complexity of magnetism may lead to numerous fascinating magnetotransport properties resulting from the magnetization-dependent anomalous Hall conductivity(AHC)~\cite{zeng2006linear}, and the establishment of a systematic and quantitative methodology for studying magnetotransport in magnetic materials will greatly facilitate the understanding of the connection between magnetoresistance and AHC.

For a quantitative analysis of the influence brought by AHE, calculations of ordinary conductivity and anomalous Hall conductivity based on a first-principles calculated tight-binding (TB) Hamiltonian are needed. Zhang \textit{et al.}~\cite{zhang2019magnetoresistance} have systematically studied the MR derived from the ordinary conductivity tensor using the Boltzmann transport theory in several realistic materials, such as Bi~\cite{zhang2019magnetoresistance}, Cu~\cite{zhang2019magnetoresistance}, SiP$_2$~\cite{MRSiP2}, $\alpha-$WP$_2$~\cite{MRalphaWP2}, MoO$_2$~\cite{MRMoO2}, TaSe$_3$~\cite{MRTaSe3}, ReO$_3$~\cite{MRReO3}, WP$_2$~\cite{zhang2019magnetoresistance}, and ZrSiS~\cite{MRZrSiS}, etc. The calculated MR curves show good agreement with experiments, within the momentum-independent relaxation time approximation. Recently, this method has been extended to study the MR and Hall effects of narrow semiconductors such as ZrTe$_5$~\cite{MRZrTe5}, and  to study the complex field, temperature and angle dependent Hall effect of non-magnetic materials~\cite{MRHallZhang2024}. Nevertheless, this approach has not been extended to magnetic materials with non-negligible AHC, where numerous magnetotransport behaviors remain to be discussed. In magnetic materials, the empirical relation for the Hall resistivity is considered to be the summation of the ordinary part $\rho_{yx}^O$ and the anomalous part $\rho_{yx}^A$, written as $\rho_{yx} = \rho_{yx}^O + \rho_{yx}^A$~\cite{Hall_1930,magnetization_1932}. Conventionally, $\rho_{yx}^O=R_o B_z$ and $\rho_{yx}^A=R_s M $, where $R_o$ is the ordinary Hall coefficient, $B_z$ is the perpendicular field and $R_s$ is the anomalous Hall coefficient.  However, this simple division of $\rho_{yx}$ may be invalid unless both the ordinary and anomalous Hall angles are small, as pointed out by Zhao \textit{et al} in Ref.~\cite{zhao2023magnetotransport}. 

Conductivity, rather than resistivity, is more essential in describing  transport physics,  considering that multiple carriers form parallel circuits contributing to the transport. Building on this perspective, our work develops a combined first-principles and semiclassical Boltzmann transport methodology to study magnetotransport in magnetic materials. We adopt the conductivity relation and extend it to multi-band magnetic materials as follows:
\begin{equation}
    {\bm \sigma}=\sum_{n}{\bm \sigma}_n^O(\bm B, T)+\bm \sigma^A(\bm B,T),\label{sigma_total}
\end{equation}
where $n$ is the band index, ${\bm \sigma}_n^O(\bm B, T)$ represent the ordinary conductivity of the $n$th band arising from Lorentz force and scattering, and $\bm \sigma^A(\bm B, T)$ represent the AHC originating from magnetism. By employing first-principles calculations and constructing a tight-binding Hamiltonian with maximally localized Wannier functions~\cite{marzari1997maximally,souza2001maximally,mostofi2008wannier90}, we can calculate the ordinary conductivity and the anomalous Hall conductivity respectively, along with the corresponding resistivity obtained by inverting the total conductivity. To gain an intuitive understanding, we initially apply this approach to several models. Typical forms of magnetism such as antiferromagnetic or paramagnetic~\cite{2023hetero,suzuki2016large,li2023field-linear,LaMnSb_2023}, soft magnetic~\cite{jaiswal2024giant,jiang2021sign}, and ferromagnetic~\cite{zhao2006fabrication,sakurai2008large,guin2019zero} behaviors are commonly observed in experiments. By considering different types of magnetization curves corresponding to these magnetic properties, a wide range of MR curves (particularly the negative MR) and Hall curves emerge, some of which match the characteristic features of several materials. We then apply our methodology to the realistic ferromagnetic material \CSS. This magnetic Weyl semimetal has drawn significant attention in recent years due to its large AHE, giant anomalous Hall angle(AHA), and unusual magnetotransport behaviors such as negative MR at high temperatures, bowtie-shaped resistance, and reverse jumping direction of MR at coercive fields at around 2K and 30K ~\cite{wang2018large_nature.commu,liu2018giant_nature.phy,shen2020Giant-anomalous-Hall_AFM,tanaka2020topological,liu2018giant_nature.phy,doi:10.1021/acs.nanolett.0c02219,zeng2021anomalous-low-resistance,zhao2023magnetoresistance,zhao2023magnetotransport}. Our calculated magnetotransport behaviors exhibit consistency with experimental observations, in terms of magnetic field and temperature dependencies. Our work establishes guidelines and provides a comprehensive and efficient approach to investigate the unusual magnetotransport properties in a wide range of magnetic materials.

\section{METHODOLOGY}\label{method}
The Boltzmann transport theory has been successfully
used for magnetoresistance calculations~\cite{zhang2019magnetoresistance}. By solving the Boltzmann equation within the framework of the relaxation time approximation, we obtain the ordinary conductivity tensor of the target bands~\cite{RGChambers_1952,ashcroft1976solid,liu2009ab,zhang2019magnetoresistance}:
\begin{gather}
    \begin{split}
        \bm \sigma_n^O(\bm B,T)&=e^2 \tau_n(T) \int_{\rm BZ}\frac{\dif \bm k}{(2\pi)^3}(-\frac{\partial f(\epsilon,T)}{\partial \epsilon})_{\epsilon=\epsilon_n(\bm k)}\\
        &\times \bm v_n(\bm k)\bar{\bm v}_n(\bm k), \label{sigma_ohe}
    \end{split}
\end{gather}
where $\tau_n(T)$ is the relaxation time of the $n$th band with temperature dependence, $f$ is the Fermi-Dirac distribution, $\bm v_n(\bm k)$ and $\bar{\bm v}_n(\bm k)$ are the velocity and weighted averaged velocity, respectively:
\begin{gather}
    \bm v_n(\bm k)=\frac{1}{\hbar}\nabla_{\bm k}\epsilon_n(\bm k),\label{velocityk}\\
    \bar{\bm v}_n(\bm k)=\int_{-\infty}^0 \frac{\dif t}{\tau_n} e^{t/\tau_n}\bm v_n(\bm k(t)),\label{velocityk_bar}
\end{gather}
and the differential equation describing the motion of $\bm k(t)$ driven by a magnetic field is approximated as
\begin{gather}
    \frac{\dif \bm k(t)}{\dif t}=-\frac{e}{\hbar}\bm v_n(\bm k(t))\times \bm B \label{kmotion}.
\end{gather}
The trajectory $\bm k(t)$ is solved by integrating Eq. (\ref{kmotion}). To obtain $\epsilon_n(\bm k)$ and $\bm v_n(\bm k)$, first-principles calculations are performed using the Vienna ab initio Simulation Package (VASP)~\cite{Kresse1996,Kresse1999} and a tight-binding model based on localized Wannier functions is constructed using the Wannier90 software package~\cite{MOSTOFI20142309}. Eqs. (\ref{sigma_ohe}-\ref{kmotion}) and Eq. (\ref{sigma_ahc_s}) below were implemented in the WannierTools software package~\cite{WU2018405}.

The intrinsic anomalous Hall conductivity can also be calculated by integrating the Berry curvature $\Omega_n^{\gamma}(\bm k)$ over the Brillouin Zone~\cite{AHC_2006}
\begin{gather}
    \begin{split}(\sigma_{\alpha\beta}^A)_s&=-\varepsilon_{\alpha\beta\gamma}\frac{e^2}{\hbar}\sum_n\int_{\rm BZ}\frac{\dif \bm k}{(2\pi)^3}f(\epsilon_n(\bm k))\Omega_n^{\gamma}(\bm k).\label{sigma_ahc_s}
    \end{split}
\end{gather}

$(\sigma_{\alpha\beta}^A)_s$ denotes the maximum intrinsic AHC when the magnetization perpendicular to the $\alpha$-$\beta$ plane is fully saturated. This magnetization may be spontaneously induced in ferromagnetic materials or induced by an external field in paramagnetic and antiferromagnetic materials. Although the AHC may exhibit a nonlinear dependence on magnetization, we adhere to the assumption that the AHC is proportional to the magnetization in magnets~\cite{zeng2006linear}. With the magnetization parallel to the magnetic field along the $\hat{z}$ direction, one can write the AHC as
\begin{gather}
    \sigma_{xy}^A(B,T)=\frac{M(B,T)}{M_s}(\sigma_{xy}^A)_{s},\label{sigmaA}
\end{gather}
where $M(B,T)$ and $M_s$ denote the field- and temperature-dependent magnetization and the saturated magnetization, respectively.

With the ordinary and anomalous parts of conductivity determined, we utilize Eq. (\ref{sigma_total}) to calculate the resistivity $\bm \rho=\bm \sigma^{-1}$, followed by the analysis of MR and Hall resistivity. The effects of different scattering mechanisms are considered in the relaxation times $\tau_n(T)$.  Notably, this approach is versatile, facilitating the calculation of a wide range of magnetic materials with various  magnetization forms, resulting in diverse MR and Hall resistivity curves.

\section{model analysis}
Section \ref{method} presents a comprehensive overview of the combined first-principles and Boltzmann transport theory methodology for calculating magnetotransport properties in magnetic materials. Before applying this methodology to realistic materials, we discuss its application to several simple models to gain an intuitive understanding of exotic magnetotransport behaviors.
\subsection{Arising of the negative magnetoresistance and temperature-dependent magnetotransport in two-band models}\label{various_form}

We begin with understanding the magnetotransport of a two-band model, taking into account the AHC. The revised longitudinal and Hall resistivities can be obtained by inverting the total conductivity tensor(see Appendix \ref{sigma_set})

\begin{gather}
    \rho_{xx}
    =\frac{e\left[ (n_e \mu_e\!+\!n_h \mu_h)\!+\!(n_e \mu_e\mu_h^2\!+\!n_h\mu_h\mu_e^2)B^2\right]}
    {e^2\left[(n_e\mu_e \!+\! n_h\mu_h)^2+
    (n_e \!-\! n_h)^2\mu_e^2\mu_h^2B^2 \right] \!+\! D(B)}\label{rxx},\\
    \rho_{yx}
    =\frac{e \left[(n_h\mu_h^2 \!-\! \!n_e\mu_e^2\!)B
    \!+\! \mu_e^2\mu_h^2(n_h \!-\! n_e)B^3\right] \!+\! C(B)\sigma_{xy}^A}
    {e^2\left[ (n_e\mu_e \!+\! n_h\mu_h)^2 \!+\!
    (n_e \!-\! n_h)^2\mu_e^2\mu_h^2B^2\right] \!+\! D(B)},\label{rxy}
\end{gather}
where $C(B)\!=\!(1\!+\!\mu_e^2B^2)(1\!+\!\mu_h^2B^2)$ and $D(B)\!=\!2e\left[ (\mu_h^2n_h\!-\!\mu_e^2n_e)B\!+\!\mu_e^2\mu_h^2(n_h\!-\!n_e)B^3\right]\sigma_{xy}^A\!+\!C(B){\sigma^A_{xy}}^2$ are the extra terms compared to the conventional two-band model, in which MR cannot be negative without the inclusion of $\sigma_{xy}^A$.  In contrast, negative MR becomes attainable in our revised form of $\rho_{xx}$ in Eq. (\ref{rxx}). 

Achieving NMR hinges on the competition between the increase in field-dependent terms in the numerator and denominator of $\rho_{xx}$. In the conventional two-band Drude model, the denominator always increases more slowly with the magnetic field than the numerator, resulting in positive MR. However,  in our revised two-band model, the term $D(B)$ can yield a significant additional increase with the field, leading to a greater growth in the denominator compared to the numerator, thereby facilitating NMR. To further illustrate this, note that in the denominator of $\rho_{xx}$, $\sigma_{xx}^2(0)$ appears as $e^2(n_e\mu_e+n_h\mu_h)^2$, and by rewriting $D(B)$ as $2e\left[ (\mu_h^2n_h\!-\!\mu_e^2n_e)B\!+\!\mu_e^2\mu_h^2(n_h\!-\!n_e)B^3\right]\tan\vartheta\sigma_{xx}(0)+(1\!+\!\mu_e^2B^2)(1\!+\!\mu_h^2B^2)\tan^2\vartheta\sigma^2_{xx}(0)$ where $\tan\vartheta=\sigma_{xy}^A/\sigma_{xx}(0)$ denotes the anomalous Hall angle(AHA), we can assess the impact of $D(B)$ through the AHA. With a sufficiently large AHA, NMR is attainable under the significant influence of $D(B)$. 

For a deeper understanding, we consider the simplest case where we set $n$, $\mu$, and $m$ to be identical for both bands. It is well-known that without $\sigma_{xy}^A$ included, this case leads to a non-saturating positive MR~\cite{chambers2012electrons, pippard1989magnetoresistance, ali2014large_WTe2}. However, things change considerably when $\sigma_{xy}^A$ is taken into account. Eq. (\ref{rxx}) can now be reduced to
\begin{gather}
    \rho_{xx}=\frac{1}{\sigma_{xx}(0)}\frac{x}{1+\tan^2\vartheta x^2}, \label{simple_rhoxx}
\end{gather}
where $\tan\vartheta$ represents the AHA, and $x=1+(\mu B)^2\geq 1$. Now we consider two distinct situations to analyze the conditions under which the negative MR arises.

(a) The anomalous Hall angle is field independent. In this situation $\tan\vartheta$ is a constant and the derivative of $\rho_{xx}$ is given by
\begin{gather}
    \frac{\dif}{\dif B}\rho_{xx}=\frac{1-\tan^2\vartheta x^2}{(1+\tan^2\vartheta x^2)^2}.
\end{gather}
When $x>1/|\tan\vartheta|$, $\frac{\dif}{\dif B}\rho_{xx}<0$. It appears that $|\tan\vartheta|\geq 1$ is required for $x>1/|\tan\vartheta|$ to hold consistently, resulting in $\rho_{xx}$ exhibiting a negative field dependence across the entire range of the $B$ field. However, this condition is rarely met in the experimentally accessible materials. In fact, $n, \mu$, and $m$ do not necessarily need to be identical. For instance, if we set $n_h \neq n_e$ while maintaining other parameters the same, $\rho_{xx}$ will saturate rapidly in the absence of $\sigma_{xy}^A$, implying a suppressed growth of $\rho_{xx}$. Likewise, with the consideration of $\sigma_{xy}^A$, obtaining negative MR is also easier even with $|\tan\vartheta|<1$.\par 

(b) The anomalous Hall angle is field dependent. The positive $B$-field dependent $\tan\vartheta$ may also help to achieve negative MR. To illustrate this, we maintain the same $n$, $\mu$, and $m$ for the two bands but introduce a positive slope for $\tan\vartheta$, given by $\frac{\dif}{\dif B}\tan\vartheta=\gamma>0$. Then we have:
\begin{gather}
    \frac{\dif}{\dif B}\rho_{xx}=\frac{2\mu^2 B(1-\tan^2\vartheta x^2)-2\gamma\tan\vartheta x^3}
    {(1+\tan^2\vartheta x^2)^2},\label{drhoxx}\\
    \lim\limits_{B\rightarrow 0}\frac{\dif}{\dif B}\rho_{xx}=
    \frac{-2\gamma\tan\vartheta }{(1+\tan^2\vartheta)^2}<0.\label{limdrhoxx}
\end{gather}
From Eq. (\ref{limdrhoxx}), it is observed that as $B$ approaches zero, the resistivity becomes negatively dependent on the magnetic field. With an increasing magnetic field $B$, $\frac{\dif}{\dif B}\rho_{xx}$ may either remain negative or turn positive, depending on the values of the mobility $\mu$ and the slope $\gamma$. Consequently, MR may be negative throughout the entire field range or initially negative but gradually rise up(see Fig. \ref{fig:large_NMR}(c) and the discussion in Section \ref{large_NMR}). In cases where the magnetization is positively field dependent, such as $B$-linear magnetization, which can be observed in paramagnetic and antiferromagnetic materials, achieving negative MR may be easier with a significantly large $\gamma$. In cases of hard ferromagnets whose magnetization is almost saturated, a weak field dependence may still exist, and the slope of the magnetization increases with temperature, resulting in an increasing slope $\gamma$ with temperature. Furthermore, at higher temperatures, the value of $\mu$ decreases. The decreasing $\mu$ and increasing $\gamma$ simultaneously contribute to maintaining the negativity of $2\mu^2 B(1-\tan^2\vartheta x^2)-2\gamma\tan\vartheta x^3$ in the numerator of $\frac{\dif}{\dif B}\rho_{xx}$. As a consequence, negative MR across the entire field range is achievable at high temperatures.

\begin{figure*}[htb]
\centering
\includegraphics[width=0.85\textwidth]{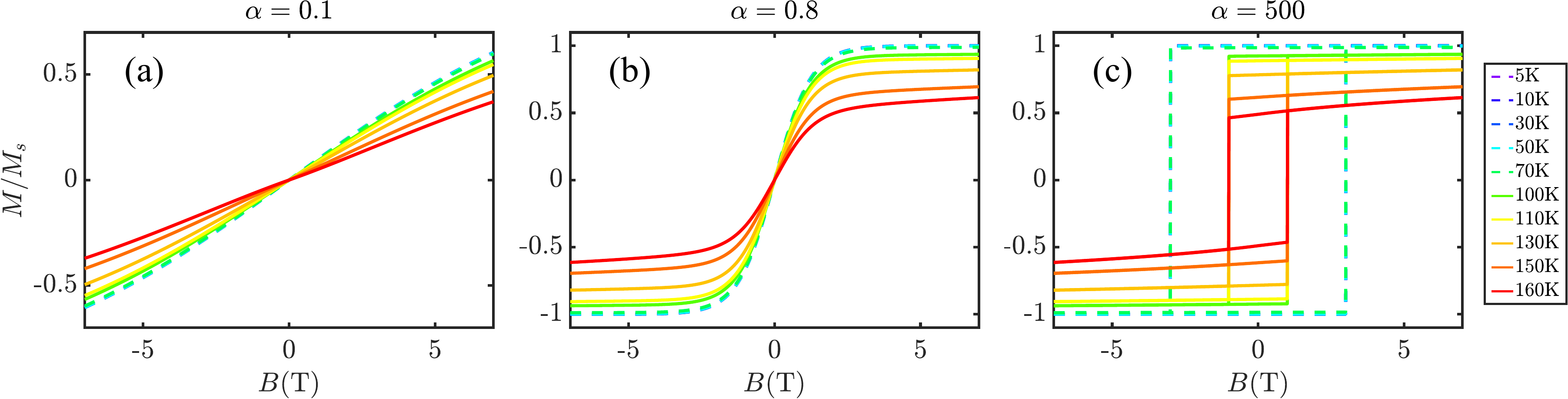}
\caption{Temperature-dependent magnetization curves, with dashed lines representing the magnetization at low temperatures and solid lines representing the magnetization at high temperatures. $\frac{M}{M_s}=\bar{S}\tanh{(\alpha B)}$ where $\alpha$ is a factor controlling the saturation speed. (a) $\alpha=0.1$. (b) $\alpha=0.8$. (c) $\alpha=500$ with a coercive field $B_c=3\rm\ T$ at low temperatures and $B_c=1\rm T$ at high temperatures.}
\label{fig:magnetization}
\end{figure*}

\begin{figure*}[htb]
\centering
\includegraphics[width=0.85\textwidth]{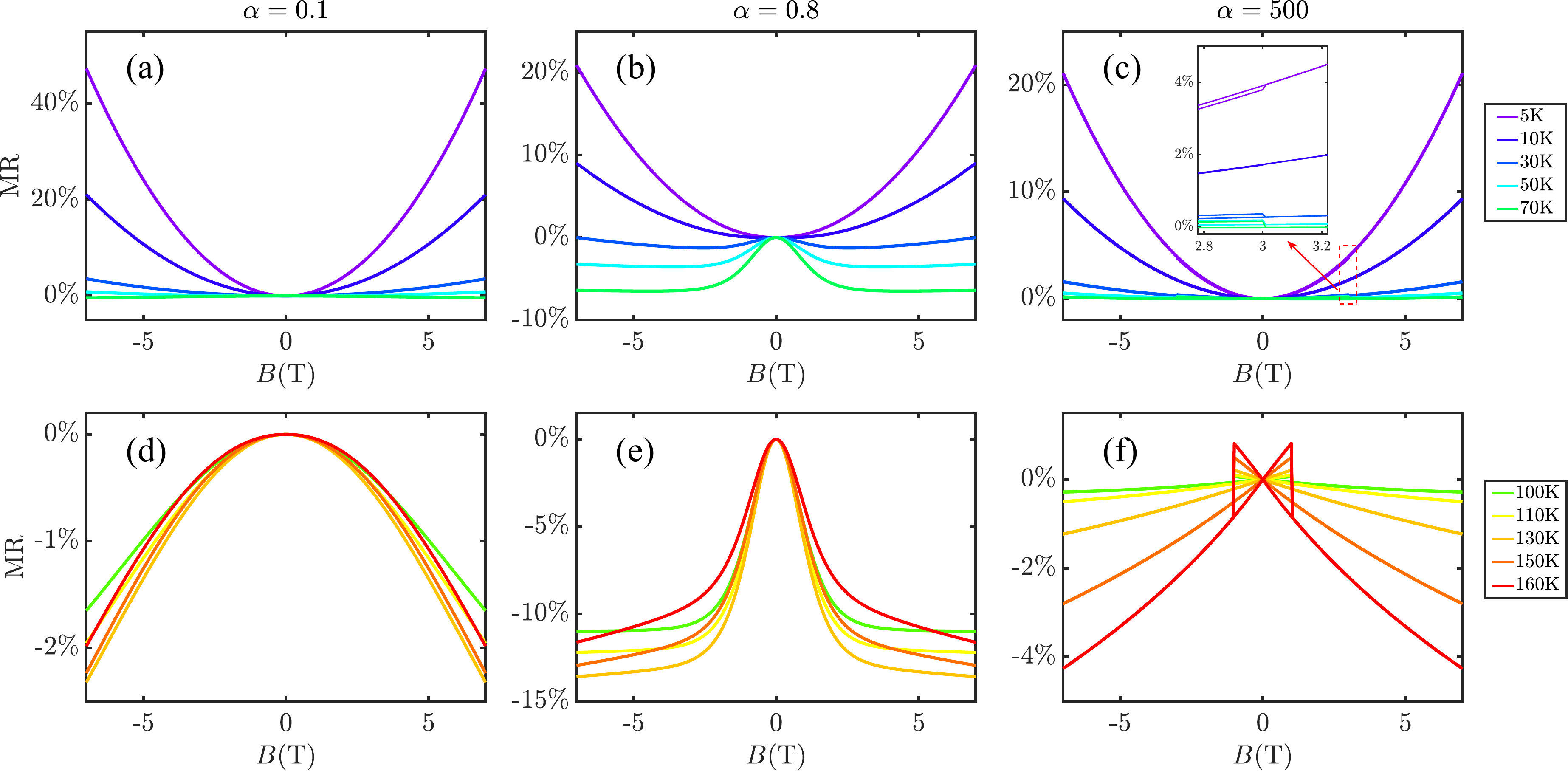}
\caption{Field- and temperature-dependent MR. (a), (d) $\alpha=0.1$, MR curves exhibit quadratic field dependence, showing positive values at low temperatures and negative values at high temperatures. (b), (e) $\alpha=0.8$, MR curves consist of two segments:  initially dropping and then rising below 70 K, or continuously dropping with different slopes above 70 K. (c), (f) $\alpha=500$, MR curves are positive and quadratic below 70 K, with the inset highlighting behaviors at $B_c=3 \rm T$. Above 70 K, MR curves turn negative and linear, displaying a bowtie shape.}
\label{fig:v2MR}
\end{figure*}

With the conditions for the emergence of NMR clarified, we now systematically analyze the evolution of MR and Hall resistivity with increasing temperature. The temperature dependence of MR and Hall resistivity is attributed to the Fermi distribution, the temperature-dependent magnetization $M(B,T)$ and mobility $\mu(T)$. Magnetization $M(B,T)$ can be generated using specific formulas or directly adopted from experimental curves for precision. Here, we generate the magnetization curves using the following formula:
\begin{gather}
    \frac{M(B,T)}{M_s}=\bar{S}\left|\tanh{(\alpha B)}\right|,\label{mag_curve}
\end{gather}
where $\bar{S}$ is the average spin determined by the mean-field method with $qJ$ controlling the Curie temperature (see Appendix \ref{Sbar} for details), and $\alpha$ is the factor controlling the saturation speed of magnetization, thereby generating different types of magnetization curves observed in experiments such as antiferromagnetic or paramagnetic, soft magnetic, and ferromagnetic behaviors.

Without loss of generality, we consider three forms of magnetization curves, corresponding to $\alpha=0.1$, $\alpha=0.8$ without a coercive field, and $\alpha=500$ with a coercive field, to represent antiferromagnetic or paramagnetic magnets, soft magnets, and ferromagnets, respectively.  These curves, depicted in Fig. \ref{fig:magnetization}, are derived from Eq. (\ref{mag_curve}) with a Curie temperature of 175K. The dashed lines and solid lines represent the magnetization at low temperatures(5-70 K) and high temperatures(100-160 K) respectively. For $\alpha=0.1$, magnetization is nearly linearly dependent on the magnetic field, as depicted in Fig. \ref{fig:magnetization}(a). For $\alpha=0.8$, magnetization grows linearly below the saturation field $B\approx 2$ T and is almost fully saturated above it, as depicted in Fig. \ref{fig:magnetization}(b). For $\alpha=500$,  magnetization is almost saturated instantaneously at first($B\approx 0$ T), as shown in Fig. \ref{fig:mag_curve}(c) in Appendix \ref{Sbar}, and by extending the field-parallel magnetization curves to the anti-parallel configuration region up to the coercive field $B_c$, with an appropriate slope, we can obtain the hysteresis magnetization loops as displayed in Fig. \ref{fig:magnetization}(c). For simplicity, we set $B_c=3$ T at low temperatures and $B_c=1$ T at high temperatures. The temperature- and field-dependent $\sigma_{xy}^A(B,T)$ is then obtainable using Eq. (\ref{sigmaA}) by setting an appropriate value for $(\sigma_{xy}^A)_{s}$. 

For the mobility $\mu(T)$, it is related to the relaxation time $\tau(T)$ by $\mu(T)=e\tau(T)/m$. In the absence of a magnetic field, according to the Drude model, $\tau(T)\propto 1/\rho(T)$~\cite{chambers2012electrons}, where $\rho(T)=\rho_0+f(T)$, $\rho_0$ is the residual resistivity, and $f(T)$ comes from the temperature-dependent scattering mechanisms such as electron-electron scattering and electron-phonon scattering, etc. For simplicity we set $\rho(T)-\rho_0\propto T$. Certainly we can add some $T^2, T^5$ scaling terms at low temperatures to improve the accuracy of the results, but they do not significantly affect the key features we care about in MR, particularly the NMR at high temperatures. Then the corresponding relaxation time has the formula $\tau(T)=1/({\rm C}+{\rm A} T)$, and the formulas for mobility can be written as
\begin{gather}
    \mu_e=\frac{e}{m_e}\frac{1}{{\rm C_e}+{\rm A_e}T},\quad \mu_h=\frac{e}{m_h}\frac{1}{{\rm C_h}+{\rm A_h}T}. \label{mobility}
\end{gather}
The parameters of Eq. (\ref{mobility}) along with the carrier concentrations and the maximum anomalous Hall conductivity are set in Table \ref{para}.

\begin{table}[h]
        \caption{parameters of our two-band models}
	\begin{tabular}{| p{3.4cm}| p{1.0cm} | p{1.0cm} |p{1.0cm}  |}\hline
	$\alpha$&$0.1$&$0.8$&$500$\\\hline
	$e/m_e(10^{11}{\rm C\cdot kg^{-1}})$&$3$&$2$&$2$\\\hline
	$e/m_h(10^{11}{\rm C\cdot kg^{-1}})$&$3$&$2$&$2$\\\hline
        ${\rm C_e}({\rm ps^{-1}})$&$1.4$&$1.4$&$1.4$\\\hline
        ${\rm C_h}({\rm ps^{-1}})$&$1.7$&$1.7$&$1.7$\\\hline
        ${\rm A_e}({\rm ps^{-1}K^{-1}})$&$0.303$&$0.303$&$0.303$\\\hline
	${\rm A_h}({\rm ps^{-1}K^{-1}})$&$0.300$&$0.300$&$0.300$\\\hline
        $n_ee(10^7\rm C/m^3)$&3.6&3.6&3.6\\\hline
        $n_he(10^7\rm C/m^3)$&4.0&4.0&4.0\\\hline
        $(\sigma_{xy}^A)_{s}(\Omega^{-1} \rm cm^{-1})$&180&180&180\\\hline
        $\rm AHA_{max}$(130 K)&0.16&0.40&0.38\\\hline
	\end{tabular} \label{para}
\end{table}
Using the magnetization curve and relaxation time settings described above, the temperature- and magnetic field-dependent MR curves of our two-band model are depicted in Fig. \ref{fig:v2MR}, corresponding to different types of magnetic materials.

\textbf{Antiferromagnetic or paramagnetic materials: }{\boldmath $\alpha=0.1$}. The MR curves are quadratic,  showing positive values at low temperatures and negative values at high temperatures, as depicted in Fig. \ref{fig:v2MR}(a)(d). As discussed in Section \ref{various_form}, a linearly field-dependent AHA with a significant slope $\gamma$ (as shown in Fig. \ref{fig:magnetization}(a)) is beneficial for achieving NMR. We found that the NMR reaches its maximum at 130 K, coinciding with the maximum value of AHA, suggesting that the enhanced AHA is responsible for the transition of MR from positive to negative.  This MR behavior, transitioning from positive to negative, is also observed in the antiferromagnetic material $\rm La Mn_xSb_2$, where $\rm Mn$ occupancy $\rm x$ controls the strength of magnetization and thus the AHA~\cite{LaMnSb_2023}. 

\textbf{Soft magnets: }{\boldmath $\alpha=0.8$}. The MR curves consist of two segments. At low temperatures, they initially decrease in the first segment (almost invisible at extremely low temperatures like 5 K and 10 K), followed by a gradual increase in the second segment, as depicted in Fig. \ref{fig:v2MR}(b). At high temperatures, MR curves exhibit a rapid decrease in the first segment followed by a slower decrease in the second segment, as illustrated in Fig. \ref{fig:v2MR}(e). The distinct MR behaviors in the two segments stem from the growth patterns of magnetization. The first segment aligns with the rapidly increasing magnetization which leads to a quick AHA rise and thus enhances the negative slope of NMR. The second segment corresponds to the saturated magnetization that indicates a relatively stable AHA, lessening its impact on the negative slope of MR. This type of NMR resembles that found in materials such as $\rm SrIrO_3$~\cite{jaiswal2024giant} and $\rm Ni_{0.8}\rm Fe_{0.2}$~\cite{jiang2021sign}, suggesting that our proposed mechanism may underlie these observations.

\textbf{Ferromagnets with magnetic coercivity: }{\boldmath$\alpha=500$}. The hysteresis magnetization loops exhibit a rectangular-like shape at low temperatures and a rhomboid-like shape at high temperatures, as depicted in Fig. \ref{fig:magnetization}(c). At high temperatures, with a slight positive slope $\gamma$ and a large AHA in the hysteresis loops, the MR curves display negative and linear field dependence with a bowtie shape, as shown in Fig. \ref{fig:v2MR}(f). At low temperatures, MR curves are quadratic and positive, and furthermore, they jump at the coercive field $B_c$ at 5 K but in contrast, drop at $B_c$ at 30 K, as shown in Fig. \ref{fig:v2MR}(c). All these behaviors are consistent with those observed in \CSS~\cite{doi:10.1021/acs.nanolett.0c02219}. The underlying reason for the reverse transition direction of the MR at the coercive field $B_c$ at 5 K and 30 K can be elucidated by the expression for $D(B)$ in the denominator of Eq. (\ref{rxx}). At $B_c$, the sign change of $\sigma_{xy}^A$ induces the sign change of $2e\left[ (\mu_h^2n_h-\mu_e^2n_e)B+\mu_e^2\mu_h^2(n_h-n_e)B^3\right]\sigma_{xy}^A$. 
Since the mobility is much less than $1 \rm{m^2/(V\cdot s)}$ (as observed in \CSS~\cite{zeng2021anomalous-low-resistance}), $(\mu_h^2n_h-\mu_e^2n_e)B\gg \mu_e^2\mu_h^2(n_h-n_e)B^3$, yielding $D(B)\approx 2e(\mu_h^2n_h-\mu_e^2n_e)B\sigma_{xy}^A +C(B){\sigma_{xy}^A}^2$. With the mobility settings provided in Table \ref{para} for $\alpha=500$ at 5 K, $(\mu_h^2n_h-\mu_e^2n_e)B\sigma_{xy}^A$ is positive when $M$ and $B$ are antiparallel, and negative when they are aligned parallelly, resulting in a decrease in the denominator of $\rho_{xx}$ and a jump in MR at $B_c$. Conversely, at 30 K, $(\mu_h^2n_h-\mu_e^2n_e)B\sigma_{xy}^A$ is negative when $M$ and $B$ are antiparallel, and positive when they are aligned parallelly, resulting in an increase in the denominator of $\rho_{xx}$ and a drop in MR at $B_c$. This suggests that the temperature dependence of mobility, which may cause the sign change of $(\mu_h^2n_h-\mu_e^2n_e)$, potentially contributes to the reverse transition direction of MR at the coercive field at different temperatures.

\subsection{Temperature-dependent magnetotransport in single-band models}\label{single_band}

\begin{figure*}
    \centering
    \includegraphics[width=0.85\textwidth]{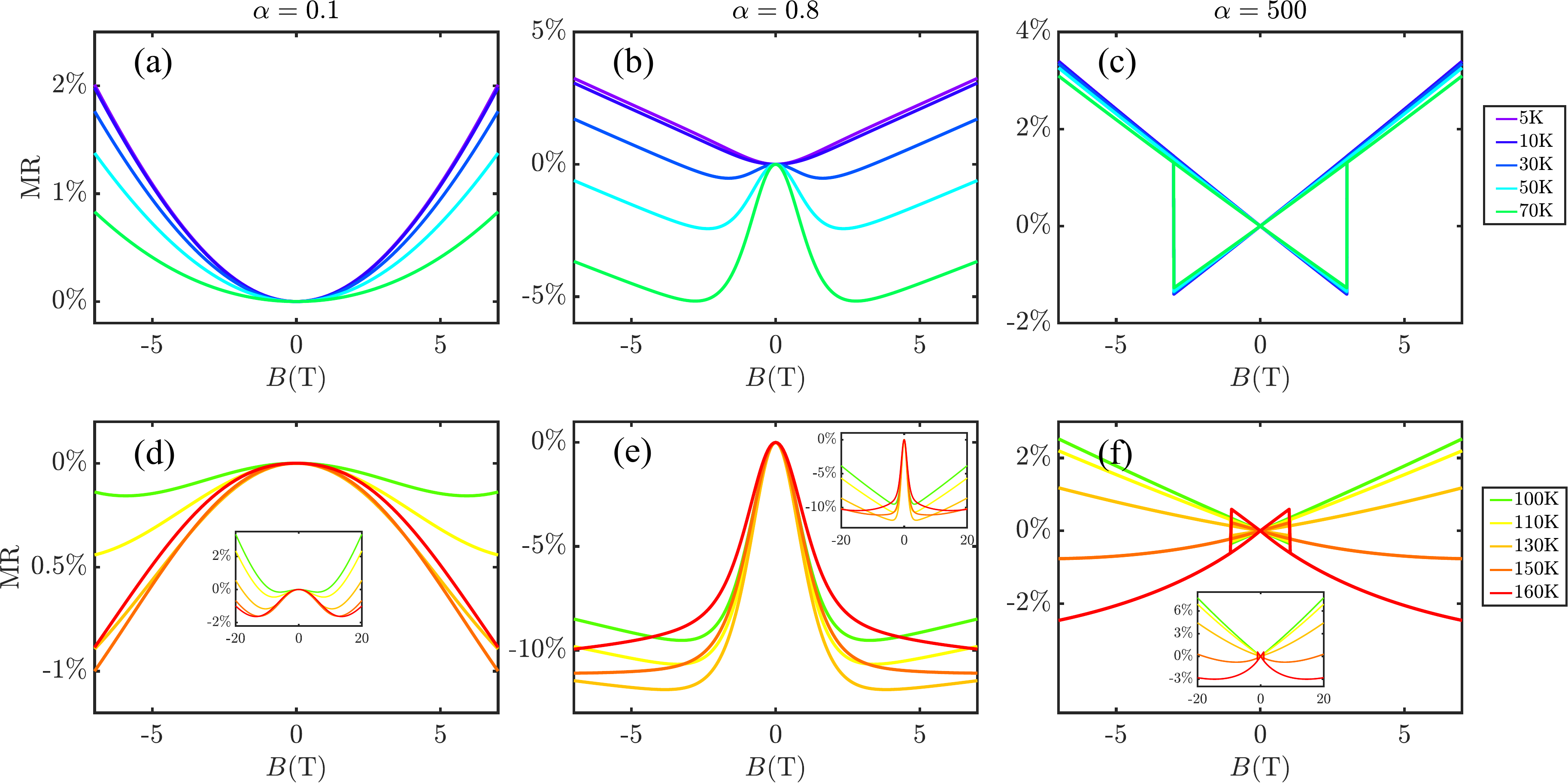}
    \caption{Field- and temperature-dependent MR curves of the electron-type single band. (a), (d) $\alpha=0.1$. MR curves are quadratic and positive below 70 K, while negative above 100K. In a wider field range, they rise and turn positive again, as shown in the inset of (d). (b), (e) $\alpha=0.8$. MR curves consist of two segments. In the first segment, they all drop down(the first segment is almost invisible at extremely low temperatures). In the second segment, they rise below 130K or initially drop down above 150 K but rise again in a wider field range, as shown in the inset of (e). (c), (f) $\alpha=500$. The MR curves are positive with a bowtie shape and linear characteristic below 130 K. With increasing temperature above 150 K, the bowtie MR curves become negative but rise again in a wider field range, as shown in the inset of (f).  }
    \label{fig:single_ele}
    \centering
    \includegraphics[width=0.85\textwidth]{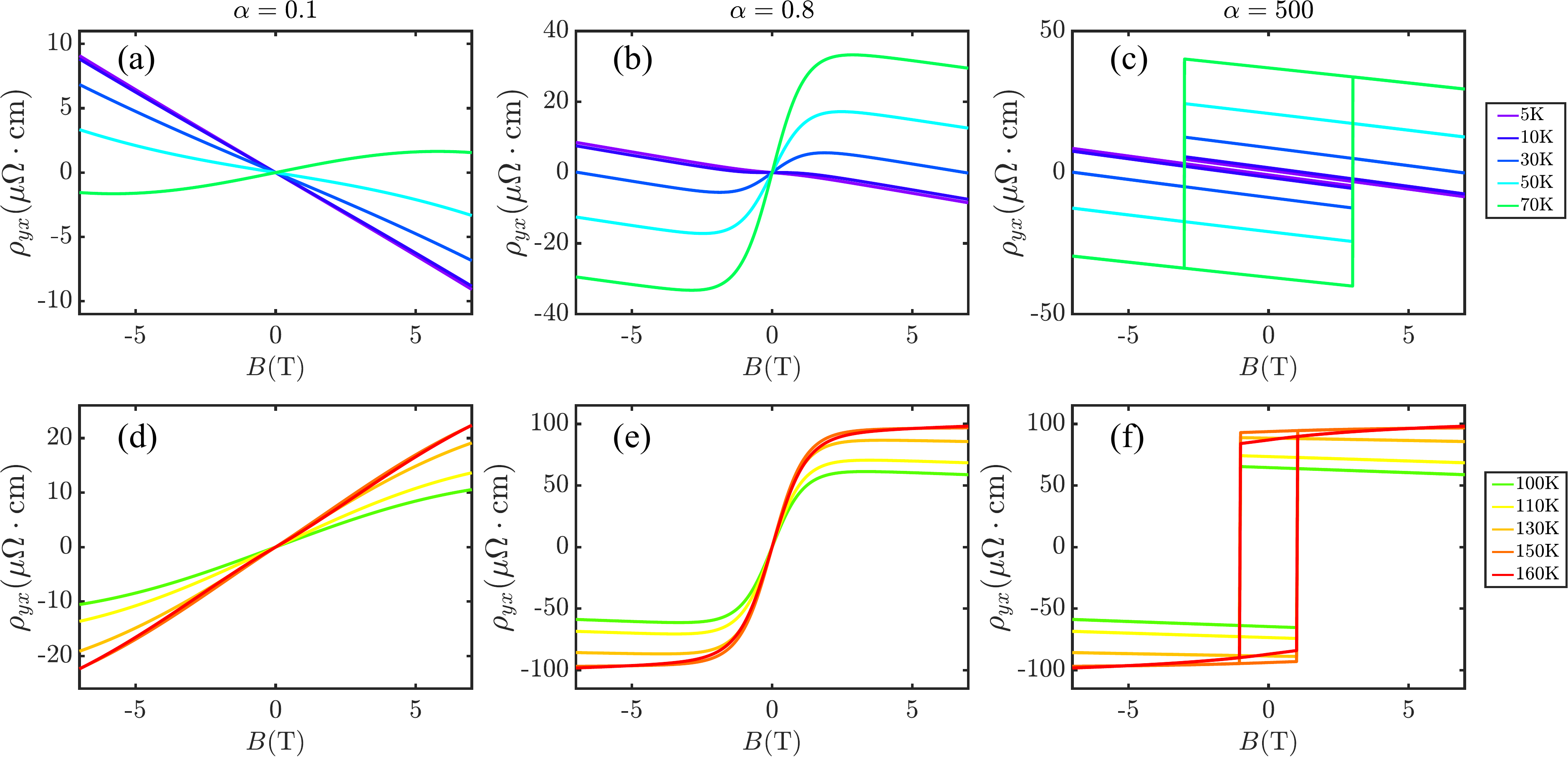}
    \caption{Field- and temperature-dependent Hall curves of the electron-type single band. (a), (d) $\alpha=0.1$. The sign reversal of the Hall resistivity is obtained by increasing the temperature without a sign change of the anomalous Hall conductivity. (b), (e) $\alpha=0.8$. The Hall resistivity also shows a transition from negative to positive field dependence. (c), (f) $\alpha=500$. The Hall curves form hysteresis loops.}
    \label{fig:rhoyx_single_ele}
\end{figure*}
In addition to the multi-band cases, single-band cases also deserve considerable discussions. The single-band conductivity is expressed as~\cite{zhao2023magnetotransport}
\begin{gather}
    \bm \sigma=\begin{bmatrix}
        \sigma_{xx}^O & \sigma_{xy}^O+\sigma_{xy}^A\\
        \sigma_{yx}^O-\sigma_{xy}^A & \sigma_{yy}^O
    \end{bmatrix},
\end{gather}
where $\sigma_{xx}^O=\sigma_{yy}^O=\sigma_0/(1+\tan^2\theta)$, $\sigma_{xy}^O=-\sigma_{yx}^O=\sigma_0 \tan\theta/(1+\tan^2\theta)$, and
the corresponding longitudinal resistivity is given by 
\begin{gather}
    \rho_{xx}=\frac{\rho_0}{(1+\tan\theta \tan\vartheta )^2+\tan^2\vartheta},\label{rhoxx_single}
\end{gather}
where $\tan\theta=\mu B$ and $\tan\vartheta=\sigma_{xy}^A/\sigma_{xx}(0)$. It is important to mention that $\tan\theta=\mu B$ corresponds to the hole-type single band, while for the electron-type single band, $\tan\theta=-\mu B$. Specifically, this means that $\tan\theta>0$ for the hole-type band and $\tan\theta<0$ for the electron-type band when $B>0$. Meanwhile, $\tan\vartheta$ assumes either a positive or negative sign depending on the specific material when $B>0$ and the magnetization is parallel to $\bm B$. Consequently, there are a total of four cases with different signs of $\tan\theta$ and $\tan\vartheta$, but their product $\tan\theta \tan\vartheta$ falls into either of two categories: positive or negative, as illustrated in Table \ref{sign}. Thus, the discussion of $\rho_{xx}$ is categorized into two cases, $\tan\theta \tan\vartheta<0$ or $\tan\theta \tan\vartheta>0$. For convenience, we keep the sign of $\tan\vartheta$ but change the sign of $\tan\theta$ according to the band type. 

\begin{table}[!htb]
        \caption{Four different combinations of signs for the ordinary Hall angle $\tan\theta$ and the anomalous Hall angle $\tan\vartheta$ when $B>0$ and the magnetization is parallel to $\bm B$.}
	\begin{tabular}{| c |>{\centering}p{1.2cm}|>{\centering}p{1.2cm} |>{\centering}p{1.2cm}|p{1.2cm} <{\centering}|}\hline
        & hole & hole& electron & electron\\ \hline
	$\tan\theta$ & $+$ & $+$ & $-$ & $-$ \\ \hline
	$\tan\vartheta$ & $+$ & $-$ & $+$ & $-$ \\ \hline
	$\tan\theta\tan\vartheta$ & $+$ & $-$ & $-$ & $+$ \\ \hline
	\end{tabular} \label{sign}
\end{table}

Linear and constant magnetization curves without temperature dependence have been employed in the single-band model introduced in Ref.\cite{zhao2023magnetotransport}. Here, we take into account the temperature dependence of mobility and magnetization for the single-band model to explore previously undiscovered properties. We still set $(\sigma_{xy}^A)_{s}$ to be $180 \ \Omega^{-1} \rm cm^{-1}$ as we did in Table \ref{para}(thus $\tan\vartheta>0$) and set $ne=7.6\times 10^7{\rm C/m^3}$ to maintain the total carrier concentration the same as that in our two-band model. For the mobility, we employ the same formula as in Eq. (\ref{mobility}), with $\mu=\frac{e}{m}\frac{1}{{\rm C}+{\rm A} T}$. $e/m$ is set to be 3, 2, 2 (in units of $10^{11}{\rm C\cdot kg^{-1}}$) for $\alpha$=0.1, 0.8, 500, respectively, and C, A are set to be $1.7{\rm ps^{-1}}$ and $0.3{\rm ps^{-1}K^{-1}}$. For the magnetization curves, we continue to use the same forms in Fig. \ref{fig:magnetization}. 

We focus on the electron-type case here due to its complexity and defer the discussion of the hole-type case in Appendix \ref{moreplots}. The calculated MR and Hall curves are plotted in Fig. \ref{fig:single_ele} and Fig. \ref{fig:rhoyx_single_ele}. In the electron-type case(corresponding to a negative $\rm tan \theta$), however, the MR may be negative at low fields(with magnetization parallel to $\bm B$), as observed in Fig. \ref{fig:single_ele} (b), (d), (e), (f), distinctly different from the positive MR at low fields mentioned in Ref.\cite{zhao2023magnetotransport}. To clarify this phenomenon we need to analyze AHA and mobility $\mu(T)$ more carefully. We start with the positive half of the magnetic field(the negative half's results are naturally obtained by symmetry). For all three forms of magnetization, the corresponding AHA at low fields can be expressed using the same formula:
\begin{gather}
    \tan \vartheta = \zeta+\gamma B, \ (0 \leq \zeta<1,\gamma>0), \label{AHA}
\end{gather}
where $\zeta=0$ for the magnetization curves of $\alpha=0.1, 0.8$ and $\zeta> 0$ for the hysteresis magnetization curves of $\alpha=500$. The longitudinal resistivity is given by $\rho_{xx}=\rho_0/[(1+\tan \theta \tan\vartheta)^2+\tan^2\vartheta]$ with $\tan\theta=-\mu B<0$ and $\tan\vartheta>0$(thus $\tan\theta\tan\vartheta<0$), and the derivative of $\rho_{xx}$ is expressed as
\begin{gather}
    \frac{\dif \rho_{xx}}{\dif B}=\frac{\rho_0 N}{[(1+\tan \theta \tan\vartheta)^2+\tan^2\vartheta]^2},\\
    \begin{split}
        N&=-2(1+\tan\theta \tan\vartheta)(\gamma\tan\theta-\mu\tan\vartheta)-2\gamma\tan\vartheta\\
        &=2\mu(1-\mu B \tan\vartheta)(2\tan\vartheta-\zeta)-2\gamma \tan\vartheta.\\
        &=2B(2\mu\gamma-\gamma^2-\mu^2\zeta^2)+2\zeta(\mu-\gamma)+O,\label{numerator}
    \end{split}
\end{gather}
where $O$ is the higher order term in $B$ expressed as $O=-6\mu^2\zeta\gamma B^2-4\mu^2\gamma^2B^3$. Dropping this non-positive term does not affect the analysis of NMR presented below. Thus, by retaining the terms of $N$ to the first order in $B$, we obtain
\begin{gather}
    N=2B(2\mu\gamma-\gamma^2-\mu^2\zeta^2)+2\zeta(\mu-\gamma).\label{numer_o1}
\end{gather}
Now we divide the discussions into two cases: $\alpha=0.1, 0.8$ with $\zeta=0$ and $\alpha=500$ with $\zeta>0$.

\textbf{Antiferromagnetic, paramagnetic materials or soft magnets: } {\boldmath $\zeta=0$}. Eq. (\ref{numer_o1}) can now be simplified to 
\begin{gather}
    N=2B\gamma(2\mu-\gamma).
\end{gather}
If $\gamma$ is large enough(determined by the shape of the magnetization curve) or $\mu$ is sufficiently small(achievable at high temperatures) such that $\gamma>2\mu$, we anticipate that $N<0$ and the MR is negative at low fields. For example, in the case of $\alpha=0.1$, the MR is positive at low temperatures as shown in Fig. \ref{fig:single_ele}(a), since $2\mu$ is larger compared to $\gamma$. However, at high temperatures with $\mu$ sufficiently small, the MR turns negative at low fields, as shown in Fig. \ref{fig:single_ele}(d). In the case of $\alpha=0.8$, $\gamma$ is large enough that negative MR is attainable even at low temperatures, as depicted in Fig. \ref{fig:single_ele}(b).

\textbf{Ferromagnets: } {\boldmath $\zeta>0$}. Scrutinizing the first term $2B(2\mu\gamma-\gamma^2-\mu^2 \zeta^2)$ of $N$ in Eq. (\ref{numer_o1}), we find it to be negative when
\begin{gather}
    \mu(T)<\frac{\gamma}{\zeta^2}(1-\sqrt{1-\zeta^2}).\label{mu_highT}
\end{gather}
Under the condition of Eq. (\ref{mu_highT}), the second term $\mu-\gamma$ of $N$ becomes
\begin{gather}
    \begin{split}
        \mu(T)-\gamma&<\frac{\gamma}{\zeta^2}(1-\sqrt{1-\zeta^2})-\gamma\\
        &=\frac{\gamma}{\zeta^2}\sqrt{1-\zeta^2}(\sqrt{1-\zeta^2}-1)<0.
    \end{split}
\end{gather}
Therefore, when the temperature is sufficiently high, satisfying Eq. (\ref{mu_highT}), $N$ is negative, leading to negative MR. In the case of $\alpha=500$ with an extremely small $\gamma$, negative MR does not appear above $B_c$ until the temperature reaches 150 K, as shown in Fig. \ref{fig:single_ele}(c), (f).

\begin{figure*}
    \centering
    \includegraphics[width=0.85\textwidth]{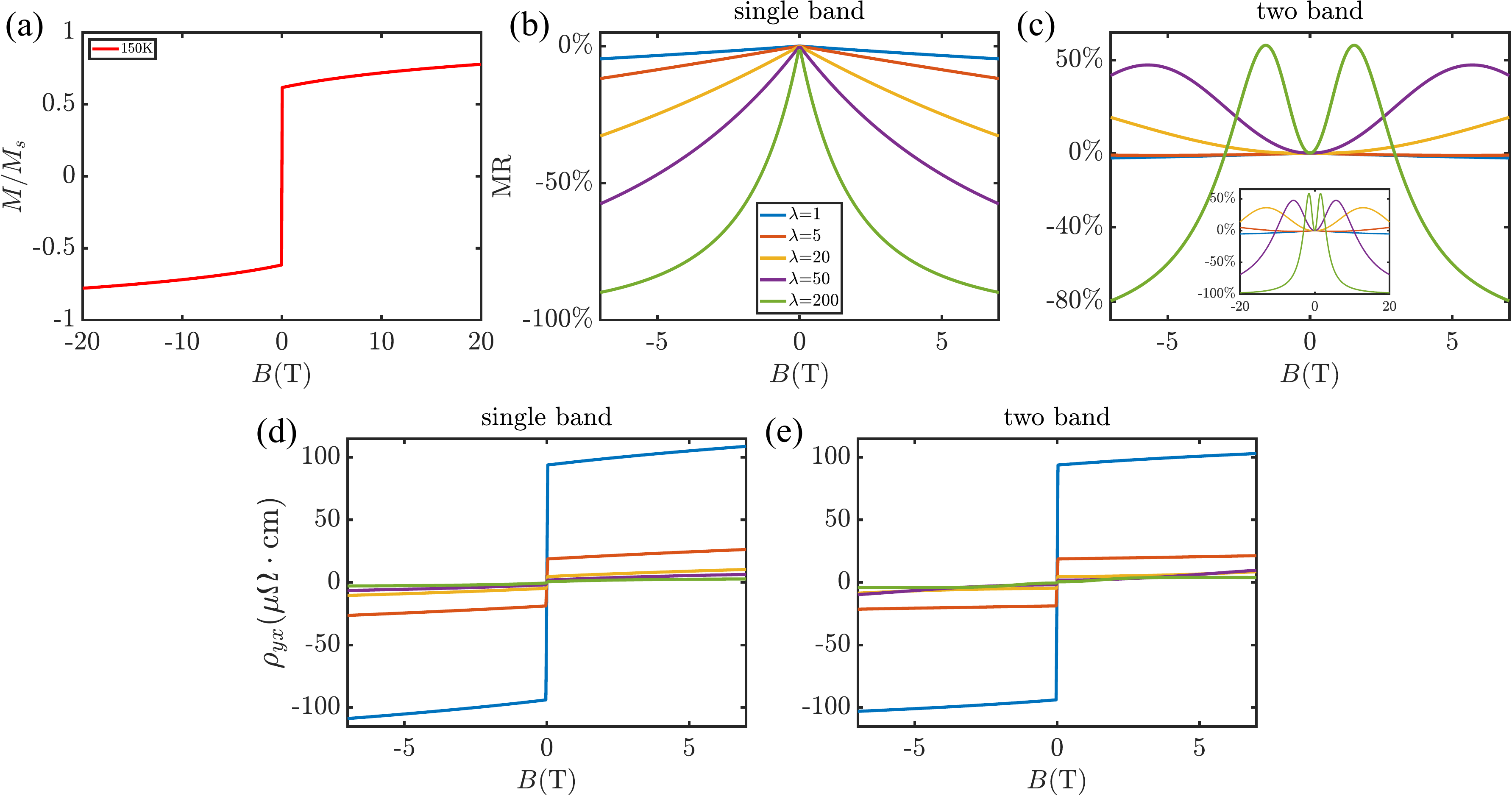}
    \caption{(a) The spontaneous magnetization without a coercive field at 150 K, which is adopted in (b)-(e). (b)-(e) The MR and Hall curves with increasing mobility while keeping the AHA unchanged(the anomalous Hall conductivity increases simultaneously with the same scaling of mobility). (b) The negative MR of the hole-type single band. (c) Various MR curves of the two-band models. Large NMR is achievable with high mobility, as shown in (b), (c). (d), (e) The Hall curves of the single-band and two-band models. Surprisingly the Hall curves are nearly identical, even though the MR curves are distinctly different.}
    \label{fig:large_NMR}
\end{figure*}

We can also analyze the derivative of the longitudinal resistivity beyond low-field regions. If the magnetization evolves slowly enough at a certain field point, we can expand the AHA to the first order in its neighborhood using the same formula in Eq. (\ref{AHA}). Here, $\zeta$ and $\gamma$ are variable, depending on the field point at which the AHA is expanded, and Eq. (\ref{numerator}) still applies. As the field increases, the magnetization gradually saturates at high fields. In this case, $\gamma$ tends to decrease to zero, and $\tan\vartheta$ approaches a constant value $\zeta_c$. Thus the asymptotic expression of Eq. (\ref{numerator}) can be reduced to
\begin{gather}
    N=2\mu\zeta_c(1-\mu B\zeta_c).
\end{gather}
$N>0$ if the mobility is not too large, and thus the derivative of $\rho_{xx}$ is positive. This explains the rising behavior of MR at sufficiently large fields, even though its field dependence is negative at low fields, as depicted in the inset of Fig. \ref{fig:single_ele} (d)-(f). The Hall curves for the electron-type case are also plotted, as shown in Fig. \ref{fig:rhoyx_single_ele}.

\subsection{Approach to obtain large negative MR} \label{large_NMR}

Large negative MR is another significant phenomenon that has garnered considerable interest~\cite{BaMnBi_2021,TbPdBi_2023}. We explore the approach to obtain large NMR in both the hole-type single-band model and two-band model in the following text. An explicit conclusion is drawn that with the AHA remaining unchanged, sufficiently large mobility facilitates the emergence of extremely large NMR.

We apply the magnetization curve of Fig. \ref{fig:magnetization}(c) at 150 K but without a coercive field, as shown in Fig. \ref{fig:large_NMR}(a), to our single-band and two-band models. Fig. \ref{fig:large_NMR}(b)-(e) depict the calculated NMR and Hall curves. The MR curves of the single-band and two-band models differ significantly, but surprisingly, their Hall curves are almost identical, indicating the same anomalous Hall nature.

\begin{figure*}[!htb]
\centering
\includegraphics[width=0.85\textwidth]{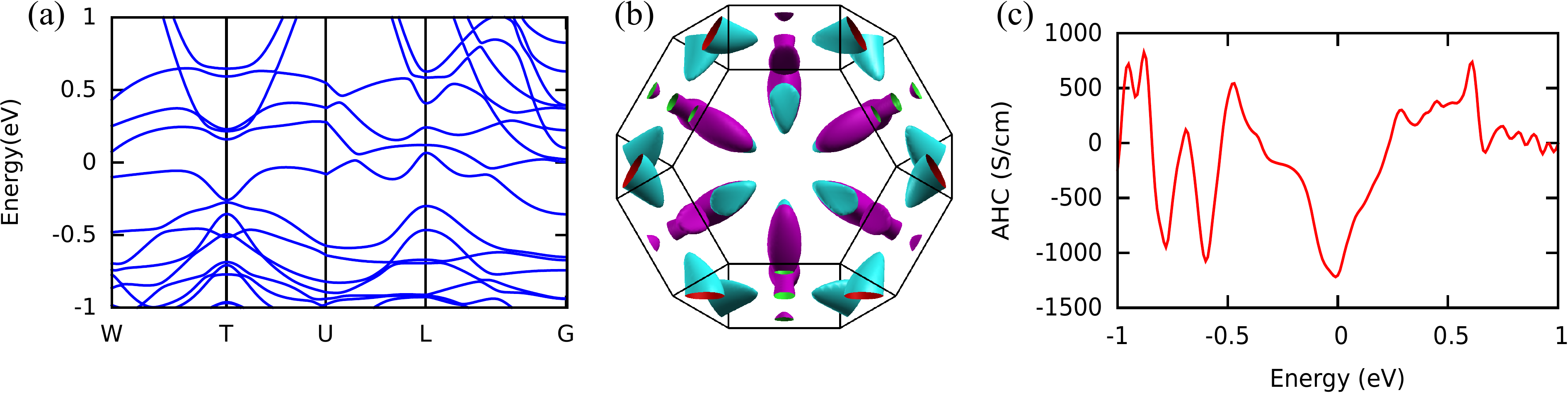}
\caption{First-principles calculations of band structure and anomalous Hall conductivity of \CSS. (a) The band structure of \CSS \ with spin-orbit coupling included. (b) Fermi surface of \CSS. The purple pocket represents the hole carrier, and the blue pocket represents the electron carrier. (c) Energy dependence of the anomalous Hall conductivity. The value at 0 eV is consistent with experimental results.}
\label{fig:band_AHE}
\end{figure*}

For the single-band model, we use the same parameters as those in Section \ref{single_band}, except for the mobility and the anomalous Hall conductivity, which are controlled by the scaling factor $\lambda$. When $\lambda=1$, the mobility and the anomalous Hall conductivity are identical to those in Section \ref{single_band}. By increasing $\lambda$, both the mobility and the anomalous Hall conductivity are multiplied by $\lambda$ simultaneously, keeping the AHA unchanged. Fig. \ref{fig:large_NMR}(b) shows the NMR for different values of $\lambda$. It is evident that the higher the mobility, the larger the NMR becomes. To illustrate this, we adopt Eq. (\ref{rhoxx_single}) and the MR can be written as: 
\begin{gather}
 \begin{split}
     \rm MR &= \frac{\rho_{xx}(B)-\rho_{xx}(0)}{\rho_{xx}(0)}\\
     &=(1+\tan^2\vartheta_0)/[(1+\tan\theta\tan\vartheta_0)^2+\tan^2\theta]-1,\label{MR_single}
 \end{split}
\end{gather}
where $\tan\vartheta$ is the AHA, with $\tan\vartheta_0$ being its value at zero field. Higher mobility results in a larger $\tan\theta$ in the denominator, leading to larger NMR. This type of large NMR resembles that observed in TbPdBi~\cite{TbPdBi_2023}. 

For the two-band model, MR curves for different values of $\lambda$ are presented in Fig. \ref{fig:large_NMR}(c). When $\lambda=1$, MR is negative and exhibits linear field dependence. When $\lambda=5$, MR is initially negative but gradually rises and becomes positive, as shown in the inset of Fig. \ref{fig:large_NMR}(c). These two cases align with the discussion in Section \ref{various_form} regarding the arsing of negative MR. For larger $\lambda$, MR is initially positive, reaches a peak, and then drops down with increasing field. An extremely large NMR approaching -100\% is achievable with $\lambda=200$ at $B=20$ T. This type of large NMR with a peak is reminiscent of observations in $\rm BaMn_2Bi_2$~\cite{BaMnBi_2021}.

To elucidate the observed peak and large NMR, we analyze MR using $\rho_{xx}$ in Eq. (\ref{simple_rhoxx}):
\begin{gather}
    {\rm MR}=\frac{\rho_{xx}(B)}{\rho_{xx}}-1=\frac{(1+\tan^2\vartheta_0)}{g(x)}-1,
\end{gather}
where $x=1+(\mu B)^2\geq 1$, $g(x)=1/x+\tan^2\vartheta x$, and $\tan\vartheta$ is the AHA with $\tan\vartheta_0$ being its value at zero field. If the growth rate of $\tan\vartheta$ is much smaller compared to $x$(which is the case with high mobility), regarding $\tan\vartheta$ as a constant is a reasonable approximation. Under this approximation we analyze the monotonicity of $1/x+\tan^2\vartheta x$ in MR:
\begin{gather}
    g'(x)=(x^2\tan^2\vartheta-1)/x^2.
\end{gather}
$g'(x)\leq 0$ when $1\leq x \leq 1/|\tan\vartheta|$ and $g'(x)>0$ when $x>1/|\tan\vartheta|$. Consequently, MR increases, encounters its peak at around $x_c=1+(\mu B_c)^2=1/|\tan\vartheta|$, and drops down thereafter, with $B_c$ being the field at which the peak is located. With larger $\lambda$(larger mobility), a smaller $B_c$ is needed, as observed in Fig. \ref{fig:large_NMR}(c). Furthermore, we notice that $g(x)$ increases rapidly with $x$ without an upper limit when $x>x_c$. Larger mobility results in a larger $x$, contributing to a larger $g(x)$, and thus extremely large NMR is obtainable.

\section{FIRST-PRINCIPLES CALCULATION OF MAGNETOTRANSPORT OF \CSS}

In Section \ref{various_form} we found that the magnetoresistance curves for the case of $\alpha=500$ are quite consistent with the observations of the realistic material \CSS. So moving beyond model analysis, we employ first-principles calculations for \CSS \ to quantitatively study its magnetotransport as our testing example. 

Recognized as a magnetic Weyl semimetal with a stacked kagome lattice structure, \CSS\ has attracted considerable attention for its large anomalous Hall angles (AHA) and unusual magnetotransport behaviors, whose mechanisms are not fully understood. In this study,  we apply our first-principles methodology to explore these magnetotransport behaviors. The band structure of \CSS, as shown in Fig. \ref{fig:band_AHE}(a), is calculated using VASP~\cite{vasp_1996} with the Perdew–Burke–Ernzerhof (PBE) exchange-correlation potential. Spin-orbit coupling(SOC) is included. The lattice constants $a$=$b$=$5.3683\rm \AA$ and $c$=$3.1783\rm \AA$ are adopted from experimental data~\cite{KASSEM2016lattice}. The positions of the interior atoms are updated by relaxing the crystal structure. The maximally localized Wannier functions for $3d$ orbitals on Co, $5p$ orbitals on Sn, and $3p$ orbitals on S are used as the basis, and the corresponding TB Hamiltonian is constructed using the Wannier90 package~\cite{Pizzi2020}. The Fermi surface of \CSS \ displayed in Fig. \ref{fig:band_AHE}(b) ensures that both the hole and electron carriers contribute to the transport. The saturated AHC, calculated when the magnetic moments are aligned with the magnetic field, is obtained using WannierTools~\cite{WU2018}, as depicted in Fig. \ref{fig:band_AHE}(c), with a notable value of  1205 S/cm at the Fermi level. This value closely approximates the findings in Ref.\cite{wang2018large_nature.commu} and the experimental value reported in Ref.\cite{liu2018giant_nature.phy}. Eq. (\ref{sigmaA}) and Eq. (\ref{mag_curve}) are employed to produce magnetization curves similar to the experimental curves~\cite{guin2019zero}, by setting $\alpha=500$ and an appropriate coercive field. The temperature-dependent relaxation times are obtained by fitting to the experimental magnetoresistance curves and Hall resistivity curves at different temperatures~\cite{zhang2019magnetoresistance}.

\begin{figure}[!htb]
\hspace{-7pt}\includegraphics[width=0.40\textwidth]{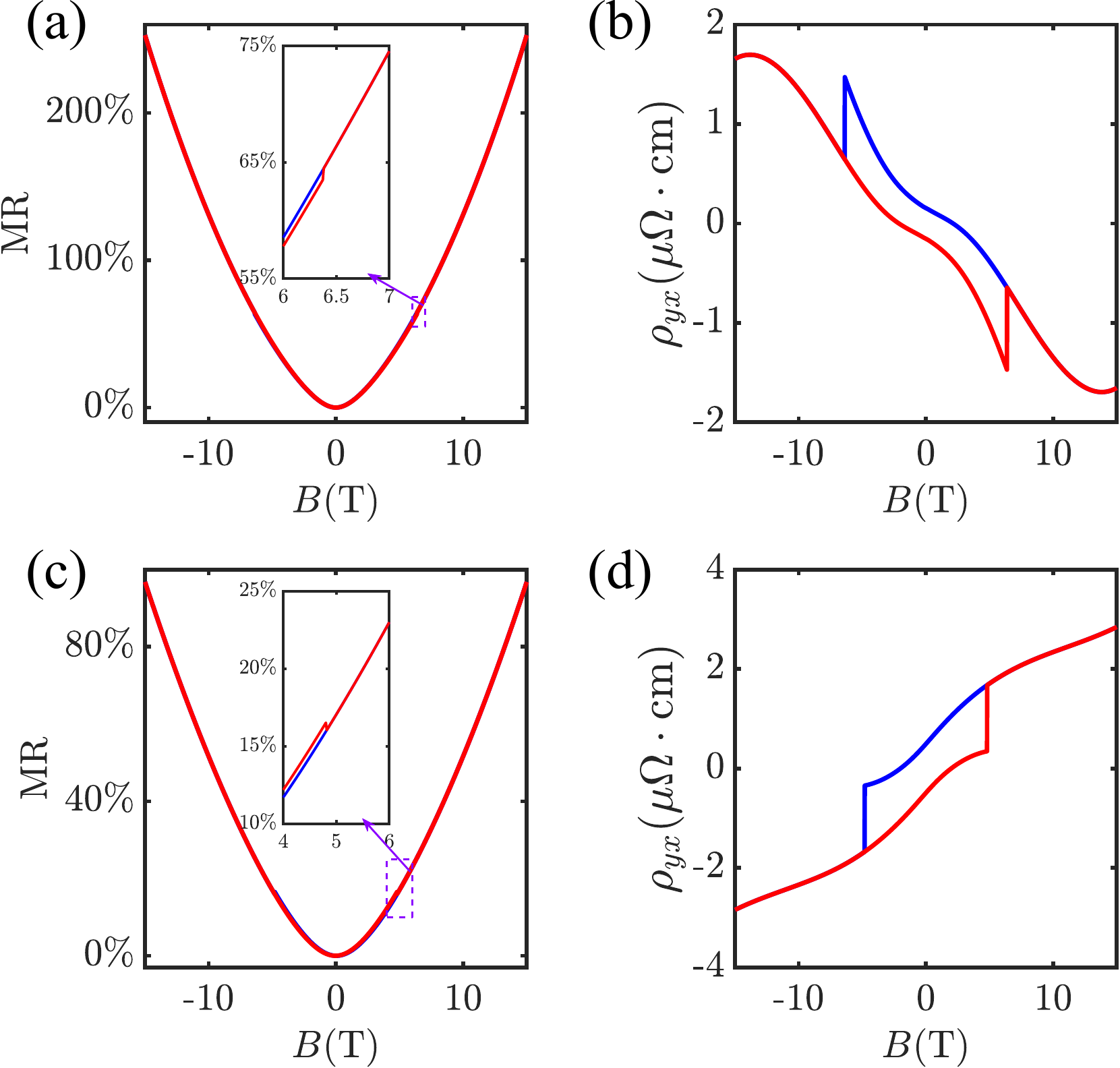}
\caption{MR and Hall resistivity calculated for \CSS \ from first-principles calculations. The insets of the MR plots show the behaviors at $B_c$. (a), (b) MR and Hall resistivity at 2 K. (c), (d) MR and Hall resistivity at 30 K.}
\label{fig:MR_lowT}
\end{figure}

\begin{figure}[!htb]
\hspace{-7pt}\includegraphics[width=0.40\textwidth]{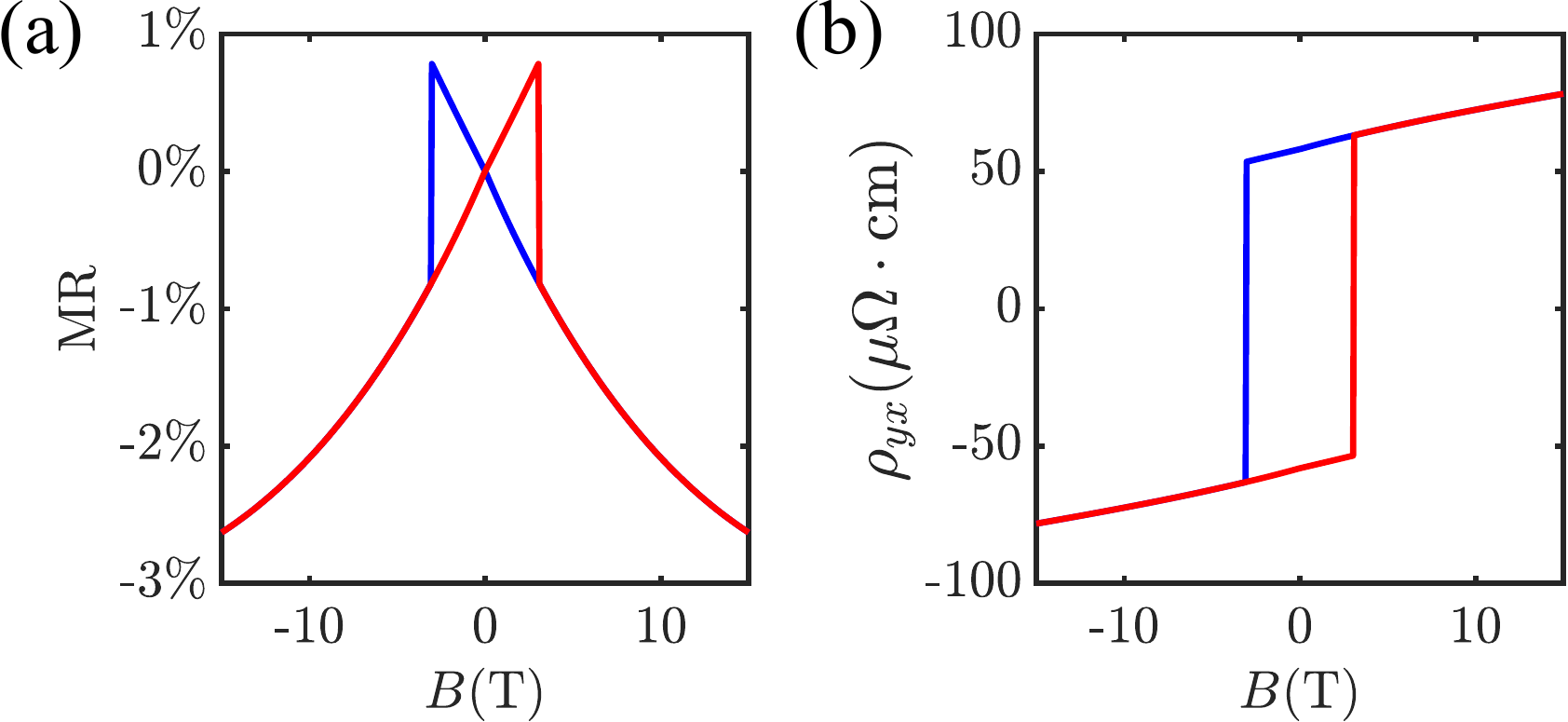}
\caption{MR and Hall resistivity calculated for \CSS \ from first-principles at 150 K. (a) MR. (b) Hall resistivity. }
\label{fig:MR_highT}
\end{figure} 

Fig. \ref{fig:MR_lowT} and Fig. \ref{fig:MR_highT} present the MR and $\rho_{yx}$ curves at 2  K, 30 K, 150 K, set with appropriate relaxation times. For comparison, the experimental curves replotted from Ref.\cite{doi:10.1021/acs.nanolett.0c02219} are shown in Appendix \ref{moreplots}. At 2 K and 30 K, the MR curves are positive and quadratic in field dependence. In contrast, at 150 K, the MR curve becomes negative and linearly field dependent with a bowtie shape, consistent with experimental results. The plots of $\rho_{xx}$ and $\sigma_{xy}$ are available in Appendix \ref{moreplots}. At 2 K, with $\tau_h=0.328 \rm ps$ and $\tau_e=0.392 \rm ps$, the MR is smaller in the magnetization configuration antiparallel to the field than in the parallel case, leading to a jump at the coercive field, as shown in Fig. \ref{fig:MR_lowT}(a). Conversely, at 30 K, with $\tau_h=0.208 \rm ps$ and $\tau_e=0.192 \rm ps$, the MR drops at the coercive field, as shown in Fig. \ref{fig:MR_lowT}(c). These properties align with the experimental results~\cite{doi:10.1021/acs.nanolett.0c02219,zeng2021anomalous-low-resistance} with no need for fine-tuning of the relaxation times. To explain these properties, we notice that at 2 K, $\tau_h<\tau_e$ but at 30 K, $\tau_h>\tau_e$, indicating that initially $\tau_e$ is larger  but decreases faster than $\tau_h$. This results in a sign change in $(\mu_h^2n_h-\mu_e^2n_e)$, as discussed in Section \ref{various_form}, accounting for the MR transitioning from a jump to a drop at $B_c$.
Regarding the calculated Hall resistivity $\rho_{yx}$ in Fig. \ref{fig:MR_lowT}(b)(d), the shapes of the hysteresis loops exhibit trends similar to those reported in  Ref.\cite{zhao2023magnetotransport}. Additionally, we want to point out that the calculated MR at 2K is larger compared to the experimental value of Ref.\cite{zhao2023magnetotransport}. This discrepancy is likely due to the absence of considerations for Berry curvature and orbital moment corrections to the ordinary conductivity tensor~\cite{xiao2005Berry,xiao2010Berry,Woo2022semi,kokkinis2022semi}, which may slow the reduction of $\sigma_{xx}^O$ with increasing field and thus suppress the growth of MR. 
\par 

At 150 K,  the relaxation time significantly decreases,  resulting in a reduced value of  the zero field longitudinal conductance. Consequently, this leads to an increase in the AHA. As mentioned in Section \ref{various_form}, a large and positive field dependent AHA helps to enlarge the negative MR. Fig. \ref{fig:MR_highT} shows the plot of MR and Hall resistivity with $\tau_h=0.022 \rm ps$ and $\tau_e=0.016 \rm ps$. The negative MR achieves a value of -2.6\% at $B=15$ T with an AHA of 0.24.  Additionally, the NMR curve exhibits significant linear characteristic. The $\rho_{yx}$ plot is also consistent with the supplementary information of Ref.\cite{doi:10.1021/acs.nanolett.0c02219} both in shape and value.  The consistency between the calculations and experiments for these unusual behaviors further confirms the validity of our approach. 

\section{conclusions}
In conclusion, we have developed a first-principles methodology to study magnetotransport in multi-band magnetic materials. This approach combines field- and temperature-dependent ordinary conductivity and anomalous Hall conductivity. Initially applied to two-band models and single-band models, we showcased a variety of MR and Hall curves with different magnetization forms, and meticulously analyzed the mechanisms underlying these unusual behaviors.  Subsequently, we applied this method to the realistic magnetic material \CSS \ using first-principles calculations. The resulting unusual magnetotransport behaviors are consistent with experimental results, indicating the validity of our calculations. Our methodology may serve as a potent tool to study the magnetotransport properties across a wide range of magnetic materials. For further research, consideration could be given to the field dependence of residual resistivity $\rho_0$ due to spin disorder scattering~\cite{Haas1968_spindiorder} or  corrections to the ordinary conductivity induced by Berry curvature and orbital moment.

\begin{acknowledgments}
This work was supported by the National Key R\&D Program of China (Grant No. 2023YFA1607400, 2022YFA1403800), the National Natural Science Foundation of China (Grant No.12274436, 11925408, 11921004), the Science Center of the National Natural Science Foundation of China (Grant No. 12188101), and  H.W. acknowledge support from the Informatization Plan of the Chinese Academy of Sciences (CASWX2021SF-0102). 
\end{acknowledgments}

\appendix
\section{model settings}
\begin{figure*}[htb]
    \includegraphics[width=0.85\textwidth]{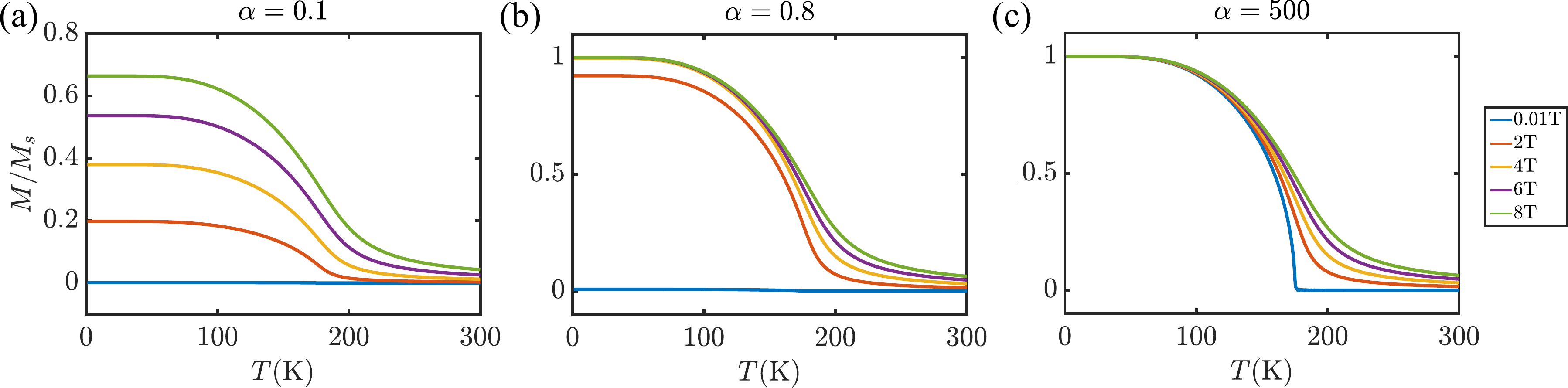}
    \caption{Magnetization curves produced by $\frac{M}{M_s}=\bar{S}\tanh{(\alpha B)}$, where $\alpha$ is the factor controlling the saturation speed. Different colors of curves represent different strength values of the magnetic field. (a) $\alpha=0.1$. (b) $\alpha=0.8$. (c) $\alpha=500$.}
    \label{fig:mag_curve}
\end{figure*}

\begin{figure*}[htb]
    \centering
    \includegraphics[width=0.85\textwidth]{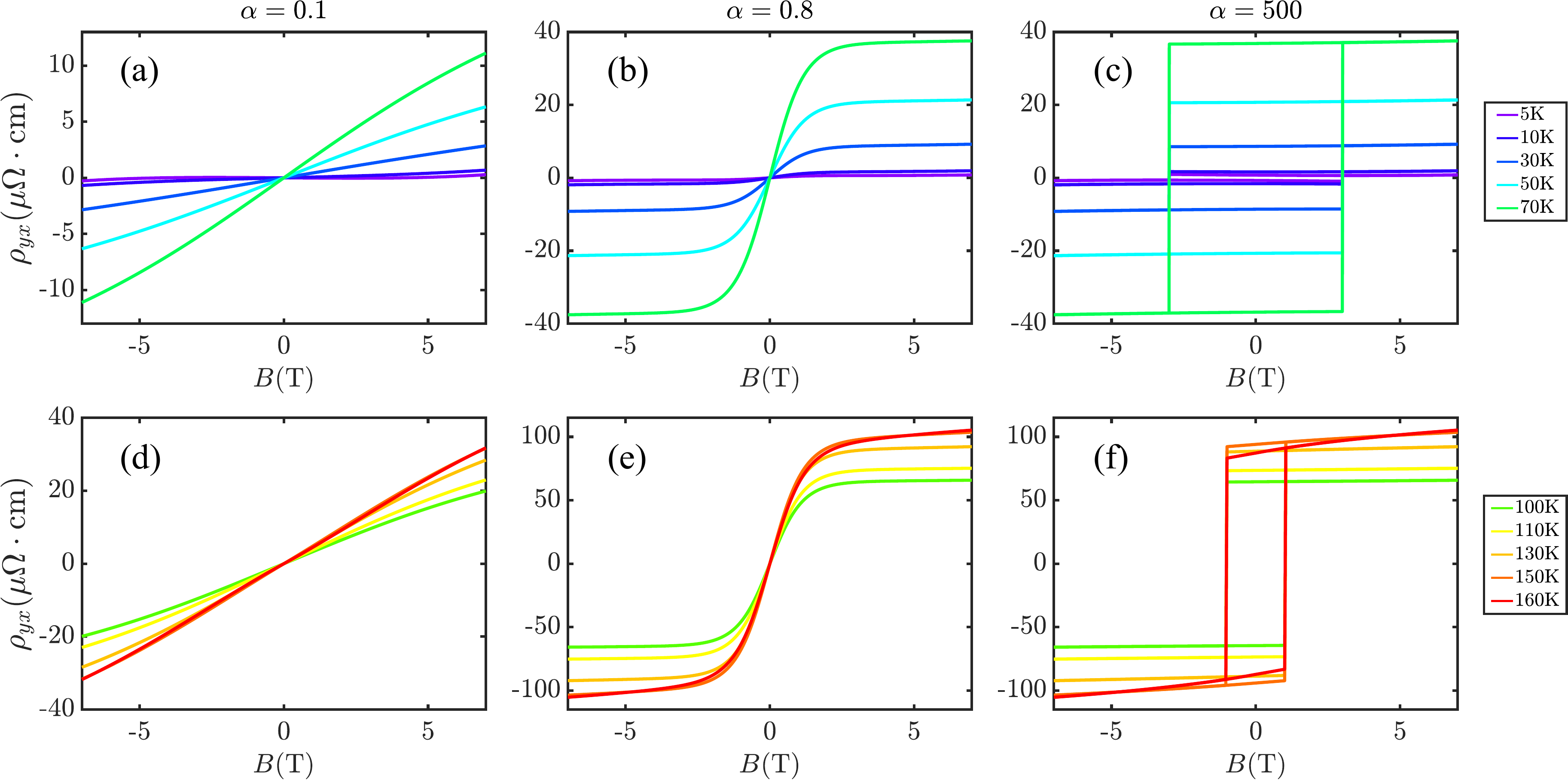}
    \caption{Field- and temperature-dependent Hall resistivity curves of our two-band model. (a), (d) $\alpha=0.1$. (b), (e) $\alpha=0.8$. (c), (f) $\alpha=500$. }
    \label{fig:v2rhoyx}
\end{figure*}

\begin{figure*}
    \centering
    \hspace{2pt}
    \includegraphics[width=0.85\textwidth]{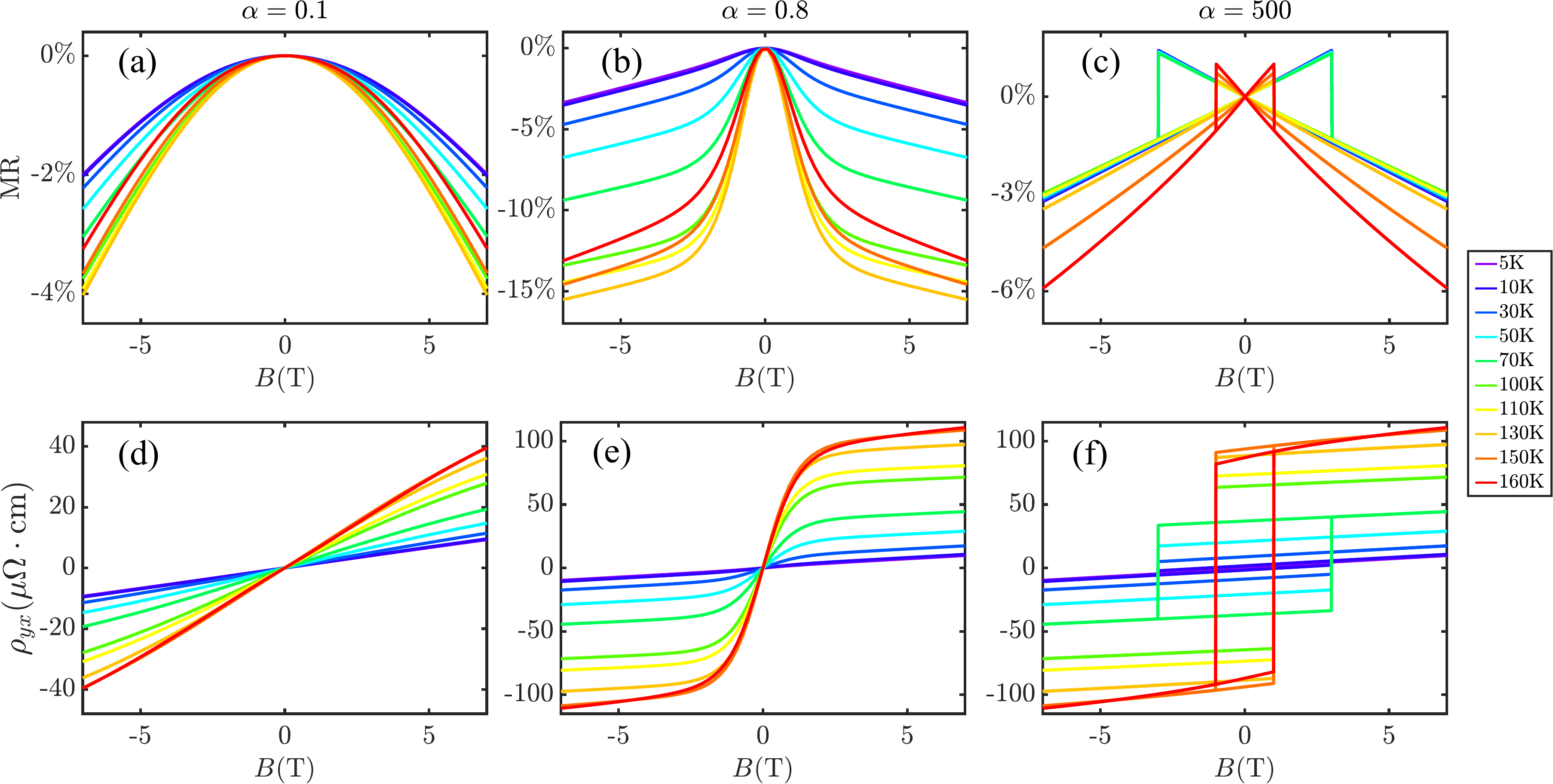}
    \caption{Field- and temperature-dependent MR and Hall curves of the hole-type single band. (a), (d) $\alpha=0.1$. The MR curves exhibit negative and quadratic field dependence, while the Hall curves display positive and linear field dependence. (b), (e) $\alpha=0.8$. The MR and Hall curves consist of two segments. The MR curves drop more rapidly in the first segment and more slowly in the second segment(the first segment is almost invisible at extremely low temperatures). The Hall curves rise rapidly in the first segment and slow down in the second segment, exhibiting behaviors similar to that of magnetization curves. (c), (f) $\alpha=500$. The MR curves exhibit negativity with a bowtie shape and linear characteristic. The Hall curves form hysteresis loops similar to that of the magnetization curves. Additionally, the Hall resistivity increases with temperature in all three cases.}
    \label{fig:single_hole}
\end{figure*}

\begin{figure*}
   \includegraphics[width=0.65\textwidth]{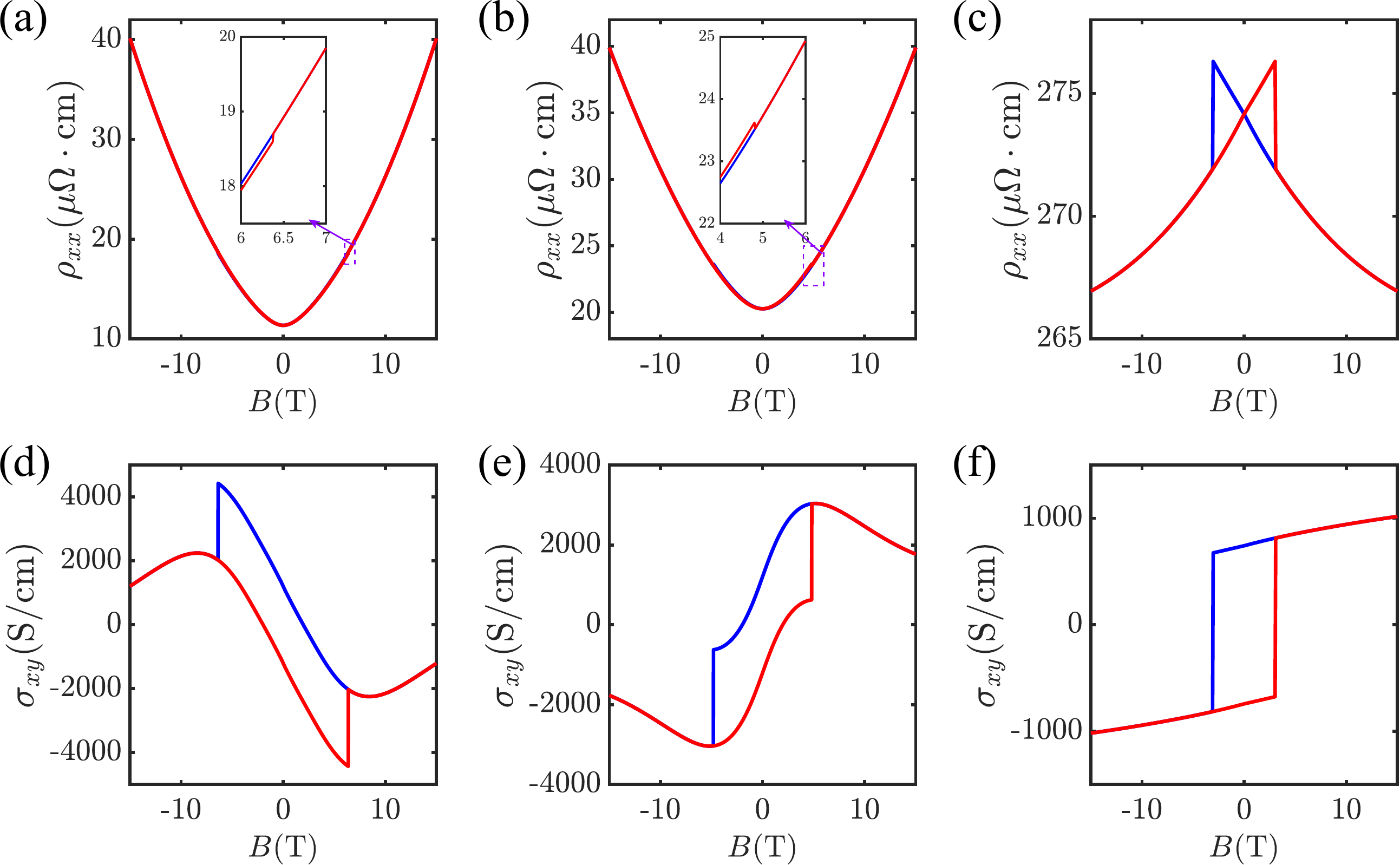}
    \caption{Longitudinal resistivity and Hall conductivity for \CSS \ from first-principles calculations. (a), (d) Longitudinal resistivity and Hall conductivity at 2 K. (b), (e) Longitudinal resistivity and Hall conductivity at 30 K. (c), (f) Longitudinal resistivity and Hall conductivity at 150 K.}
    \label{fig:plots_more}
\end{figure*}

\begin{figure*}[htb]
    \includegraphics[width=0.65\textwidth]{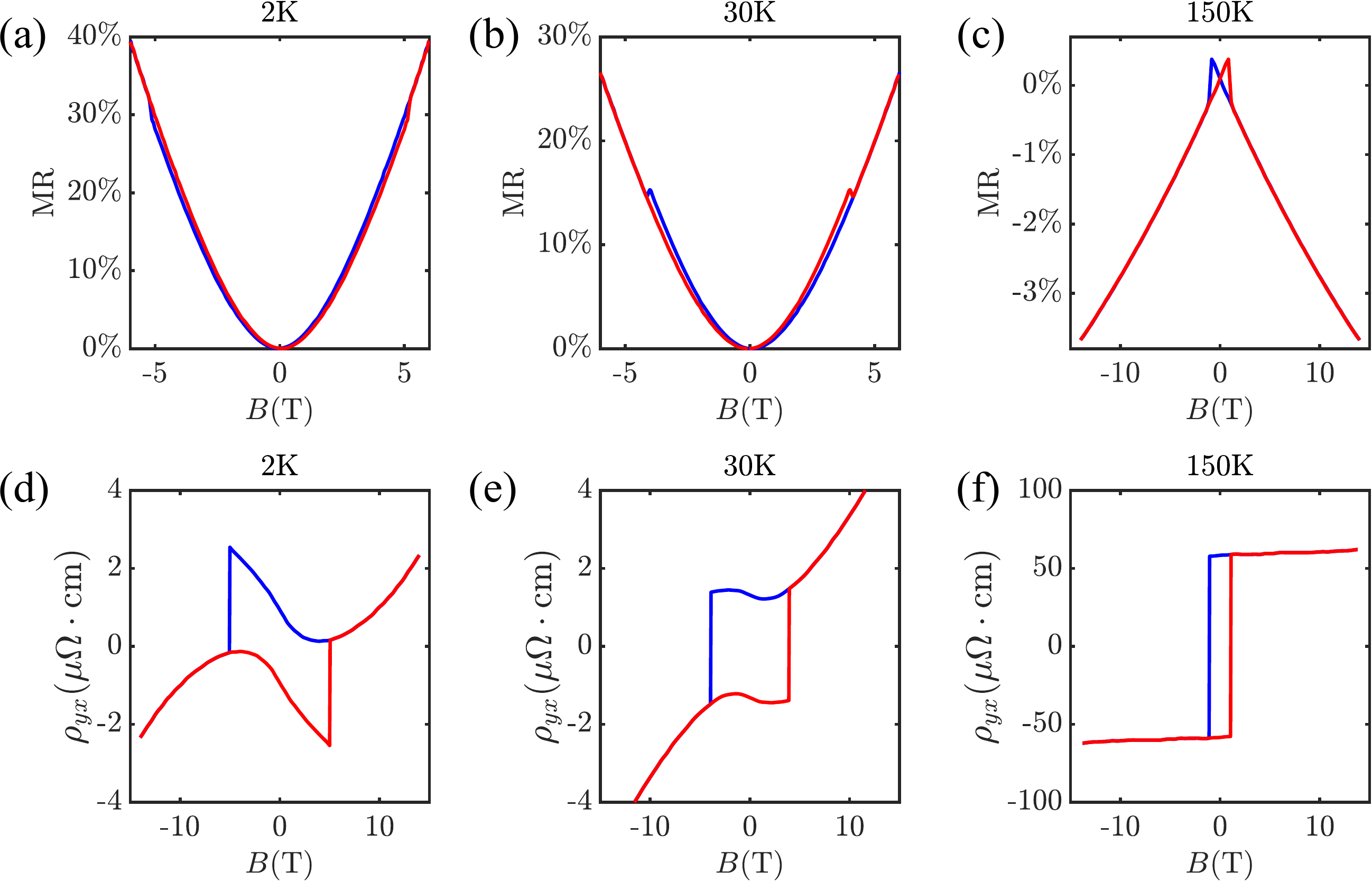}
    \caption{Experimental data of MR and Hall curves for \CSS, replotted from Ref.\cite{doi:10.1021/acs.nanolett.0c02219} under the Creative Commons Attribution(CC-BY) License. (a), (d) MR and Hall curves at 2 K. (b), (e) MR and Hall curves at 30 K. (c), (f) MR and Hall curves at 150 K}
    \label{fig:2K_30K_experi}
\end{figure*}
\subsection{Settings for $\sigma$}\label{sigma_set}

For the two-band model, the total conductivity tensor is the sum of the ordinary part ($\bm \sigma_e$, $\bm \sigma_h$) and the anomalous part $\bm \sigma^A$:
\begin{gather}
    \bm \sigma =\bm \sigma_e+\bm \sigma_h + \bm \sigma^A,\\
    \bm \sigma_e=\sigma_e^0
        \begin{bmatrix}
            \dfrac{1}{1+\mu_e^2 B^2}&\dfrac{-\mu_e B}{1+\mu_e^2 B^2}\\
            \dfrac{\mu_e B}{1+\mu_e^2 B^2}&\dfrac{1}{1+\mu_e^2 B^2}
    \end{bmatrix},\\
    \bm \sigma_h=\sigma_h^0
        \begin{bmatrix}
            \dfrac{1}{1+\mu_h^2 B^2}&\dfrac{\mu_h B}{1+\mu_h^2 B^2}\\
            \dfrac{-\mu_h B}{1+\mu_h^2 B^2}&\dfrac{1}{1+\mu_h^2 B^2}
        \end{bmatrix},\\
    \bm \sigma^A=
        \begin{bmatrix}
            0 & \sigma_{xy}^A(B,T)\\
            -\sigma_{xy}^A(B,T) & 0
        \end{bmatrix},\label{sigma_sum}
\end{gather}
where $\sigma_e^0=n_e e \mu_e$, $\sigma_h^0=n_h e \mu_h$, $e$ is the electron charge, $n_e, n_h$ are the carrier concentrations of the electron and hole type bands, $\mu_e, \mu_h$ are the mobilities of the two bands, and $\sigma_{xy}^A(B,T)$ is the field- and temperature-dependent anomalous Hall conductivity. The resistivities $\rho_{xx}$ and $\rho_{yx}$ in Eq. (\ref{rxx}) and Eq. (\ref{rxy}) of the main text are obtained by $\bm \rho=\bm \sigma^{-1}$.

\subsection{Average spin by the mean field method and the production of magnetization curves}\label{Sbar}

To simulate the magnetization curves, we have adopted the average spin of the Ising model in Eq. (\ref{mag_curve}). The Hamiltonian of the Ising model is given by
\begin{gather}
    H=-J\sum_{\langle i,j\rangle}S_i S_j-B\sum_i^NS_i, \label{Ising}
\end{gather}
where $J$ is the exchange coupling strength. Ignoring the terms representing the fluctuation of spin, $(S_i-\bar S)(S_j-\bar S)$, $S_i S_j$ is approximated as
\begin{gather}
    \begin{split}
        S_iS_j&=[\bar{S}+(S_i-\Bar{S})][\bar{S}+(S_j-\bar{S})]\\
        &=\bar S^2+\bar S(S_i-\bar S)-\bar S(S_j-\bar S)+(S_i-\bar S)(S_j-\bar S)\\
        &\approx \bar S^2+\bar S(S_i-\bar S)-\bar S(S_j-\bar S)\\
        &=-\bar S^2+\bar S(S_i+S_j).
    \end{split}
\end{gather}
Then Eq. (\ref{Ising}) can be written as
\begin{gather}
    \begin{split}
        H&=-J\sum_{\langle i,j\rangle}[-\bar S^2+\bar S(S_i+S_j)]-B\sum_i^N S_i\\
        &=\frac{1}{2}NqJ\bar S^2-\sum_i(B+qJ\bar S)S_i.
    \end{split}
\end{gather}
The single particle partition function and the total partition function are:
\begin{gather}
    Z_i=e^{-\frac{1}{2}\beta qJ\bar S^2+\beta B_{eff}}+e^{-\frac{1}{2}\beta qJ\bar S^2-\beta B_{eff}},\\
    Z=Z_1^N=e^{-\frac{1}{2}\beta NqJ\bar S^2}(2\cosh(\beta B_{eff}))^N,
\end{gather}
where $\beta=1/k_BT$, $B_{eff}=B+qJ\bar S$. The free energy is given by
\begin{gather}
    \begin{split}
        F(T,B)&=-k_BT\ln Z\\
        &=\frac{1}{2}NqJ\bar S^2-Nk_BT\ln(2\cosh(\beta(B+qJ\bar S))).
    \end{split}
\end{gather}
Finally we obtain the equation that the average spin satisfies:
\begin{gather}
    \bar S=-\frac{1}{N}(\frac{\partial F}{\partial B})_{T,N}=\tanh(\beta(B+qJ\bar S)).
\end{gather}
This equation can be solved numerically. Considering the case with $B=0$, we find that above $T_c=qJ/k_B$ the average spin $\bar{S}=0$. Thus $T_c$ represents the Curie temperature. By applying Eq. (\ref{mag_curve}) and set different values of $\alpha$, we can produce the magnetization curves as shown in Fig. \ref{fig:mag_curve}.

\section{Supplemental plots for the hole-type single band models, two-band models and \CSS\ semimetal}\label{moreplots}
\begin{figure}
    \centering
    \includegraphics[width=0.46\textwidth]{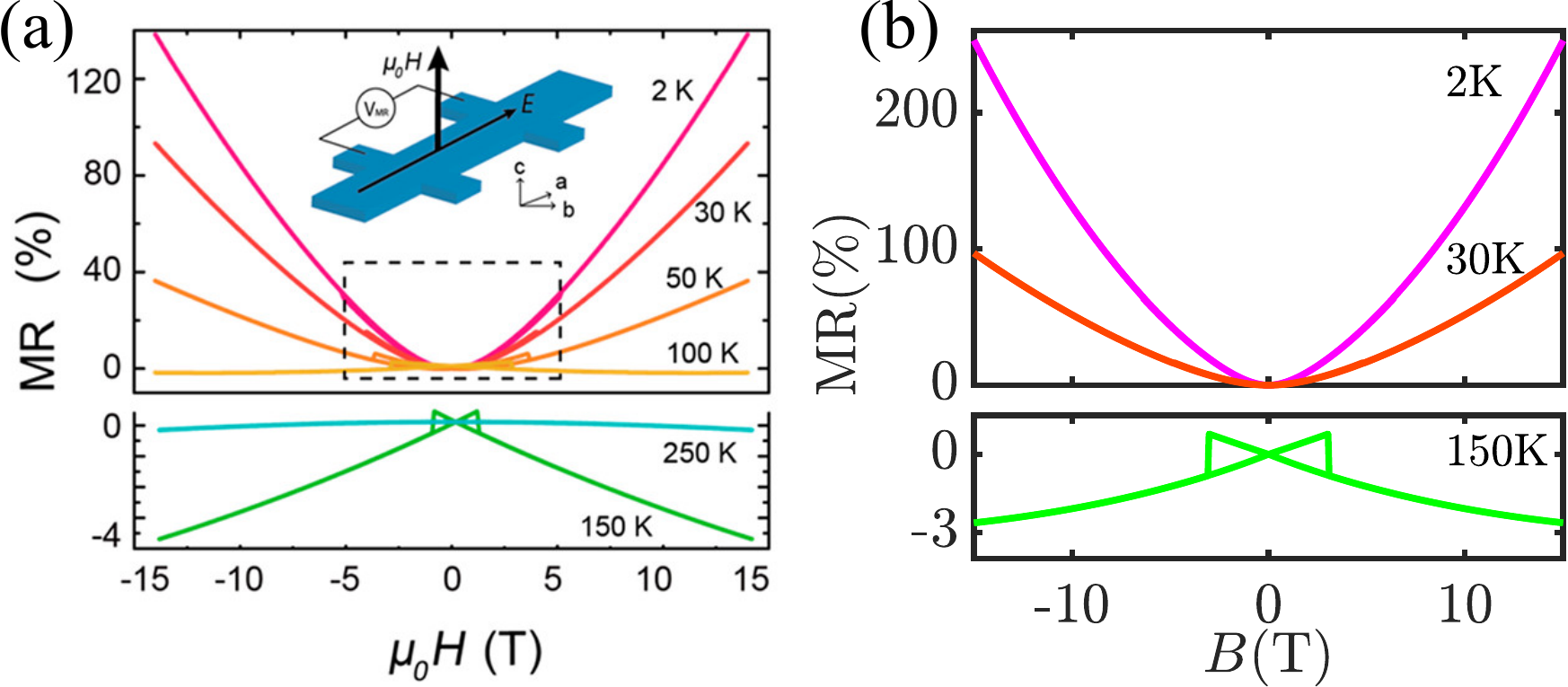}
    \caption{Comparison between experimental and first-principles calculated MR curves. (a) Experimental MR curves from 2 K to 150 K. Figure taken from Ref.\cite{doi:10.1021/acs.nanolett.0c02219} under the Creative Commons Attribution(CC-BY) License. (b) Our calculated MR curves at 2 K, 30 K and 150 K.}
    \label{fig:MR_joint_exp}
\end{figure}
In Section \ref{single_band}, we analyzed the magnetotransport of electron-type single-band models. Now we turn to the hole-type case, as shown in Fig. \ref{fig:single_hole}.
For the hole-type case(which corresponds to positive $\tan\theta$), the MR plots for linear magnetization($\alpha=0.1$) and hysteresis magnetization($\alpha=500$) are similar to Fig. 2(a) and Fig. 1(a) in Ref.\cite{zhao2023magnetotransport}, as depicted in Fig. \ref{fig:single_hole}(a), (c). In our plots, as the temperature increases, the NMR reaches its maximum strength with the maximal AHA at 130 K in the former case, whereas it continues to grow in strength up to 160 K in the latter case. As for magnetization curves with $\alpha=0.8$, the MR curves consist of two segments, as shown in Fig. \ref{fig:single_hole}(b). The curves drop more rapidly in the first segment and more slowly in the second segment. At extremely low temperatures below 10 K, where the AHA is small, the first segment is nearly invisible. Additionally, the NMR also reaches its maximum strength at 130 K. Overall, the MR shows a negative field-dependent behavior with $\tan\theta \tan\vartheta>0$, consistent with Ref.\cite{zhao2023magnetotransport}. This behavior can be illustrated using Eq. (\ref{MR_single}):
\begin{gather}
    \begin{split}
        \rm MR &=\frac{1+\tan^2\vartheta_0}{(1+\tan\theta\tan\vartheta)^2+\tan^2\theta}-1\\
        &=\frac{1-(1+\tan\theta\tan\vartheta)^2+\tan^2\vartheta_0-\tan^2\vartheta}{(1+\tan\theta\tan\vartheta)^2+\tan^2\vartheta},\label{MR_single_hole}
    \end{split}
\end{gather}
where $\tan\vartheta_0$ represents the AHA at zero field. Obviously, $\tan^2\vartheta_0-\tan^2\vartheta<0$, and given $\tan\theta\tan\vartheta>0$ with magnetization parallel to $\bm B$, we have $1-(1+\tan\theta\tan\vartheta)^2<0$. As a consequence, the MR is negative. The Hall curves exhibit similar behaviors to the magnetization curves, as shown in Fig. \ref{fig:single_hole}(d)-(f).

In Fig. \ref{fig:v2rhoyx} we present the corresponding field- and temperature-dependent Hall resistivity curves for the two-band model discussed in Section \ref{various_form}. And in Fig. \ref{fig:plots_more}(a)-(f) we present the longitudinal resistivity and Hall conductivity curves calculated for \CSS \ at 2 K, 30 K and 150 K. The experimental curves replotted from Ref.\cite{doi:10.1021/acs.nanolett.0c02219} are also shown in Fig. \ref{fig:2K_30K_experi}. For an intuitive comparison, the MR curves from experiments and first-principles calculations are placed together in Fig. \ref{fig:MR_joint_exp}.

\bibliography{refs.bib}

\end{document}